\documentclass[printer]{aa}
\usepackage{graphicx}
\usepackage[]{natbib}

\bibpunct{(}{)}{;}{a}{}{,}

\titlerunning{}

\def\kms{km~s$^{-1}$}
\def\tef{\textit{T}_{\text{eff}}}
\def\logg{\text{log}(\textit{g})}
\def\mh{[\text{M}/\text{H}]}
\def\feh{[\text{Fe}/\text{H}]}
\def\mgfe{[\text{Mg}/\text{Fe}]}
\def\alffe{[\alpha/\text{Fe}]}
\def\snr{S/N}
\def\vsini{v\sin(i)}
\def\vrad{V_{\text{rad}}}
\def\loggf{\log{gf}}
\def\ccf{\text{FWHM}_{\text{CCF}}}

\def\cms{cm~s$^{-2}$}

\begin{document}

\title{The AMBRE Project:\\\emph{r-}process elements in the Milky Way thin and thick discs
\thanks{Full \tablename~\ref{table_ambre_ba} is only available in electronic form at the 
CDS via anonymous ftp to cdsarc.u-strasbg.fr (130.79.128.5) or via http://cdsweb.u-strasbg.fr/cgi-bin/qcat?J/A+A/.}}

\titlerunning{\emph{r-}process abundances in the Milky Way thin and thick discs}
\authorrunning{Guiglion et al.}

\author{G. Guiglion \inst{1, 2}, 
P. de Laverny \inst{2}, 
A. Recio-Blanco \inst{2}, 
N. Prantzos \inst{3}}

\institute{Leibniz-Institut f\"ur Astrophysik Potsdam (AIP) An der Sternwarte 
16, 14482 Potsdam \and Universit\'e C\^ote d'Azur, Observatoire de la C\^ote d'Azur, 
CNRS, Laboratoire Lagrange, France \and Institut d'Astrophysique de Paris, UMR7095 
CNRS, Universit\'e P. \& M. Curie, 98bis Bd. Arago, 75104 Paris, France}

\date{Received 06/07/2018 ; Accepted 06/09/2018}

\abstract{The chemical evolution of neutron capture elements in the Milky Way
disc is still a matter of debate. There is a lack of statistically 
significant catalogues of such element abundances, especially  those of the 
\emph{r-}process.}{We aim to understand the chemical evolution of 
\emph{r-}process elements in Milky Way disc. We focus on three pure 
\emph{r-}process elements Eu, Gd, and Dy. We also consider a pure \emph{s-}process 
element,  Ba, in order to disentangle the different nucleosynthesis processes.}
{We take advantage of high-resolution FEROS, HARPS, and UVES spectra from the 
ESO archive in order to perform  a homogeneous analysis on $6\,500$ FGK Milky 
Way stars. The chemical analysis is performed thanks to the automatic optimization 
pipeline GAUGUIN. We present abundances of Ba (5057 stars), Eu (6268 stars), 
Gd (5431 stars), and Dy (5479 stars). Based on the $\alffe$ ratio determined 
previously by the AMBRE Project, we chemically characterize the thin and the 
thick discs, and a metal-rich $\alpha$-rich population.}{First, we find that the 
[Eu/Fe] ratio follows a continuous sequence from the thin disc to the thick disc 
as a function of the metallicity. Second, in thick disc stars, the [Eu/Ba] ratio 
is found to be constant, while the [Gd/Ba] and [Dy/Ba] ratios decrease as a function 
of the metallicity. These observations clearly indicate a different nucleosynthesis 
history in the thick disc between Eu and Gd-Dy. The [\emph{r}/Fe] ratio in the 
thin disc is roughly around +0.1 dex at solar metallicity, which is not the case for Ba.
We also find that the $\alpha$-rich 
metal-rich stars are also enriched in \emph{r-}process elements (like thick disc stars), 
but their [Ba/Fe] is very different from thick disc stars. Finally, we find that 
the $[r/\alpha]$ ratio tends to decrease with metallicity, indicating that supernovae 
of different properties probably contribute differently to the synthesis 
of \emph{r-}process elements and $\alpha$-elements.}{We  provide average abundance 
trends for [Ba/Fe] and [Eu/Fe] with rather small dispersions, and for the first 
time for [Gd/Fe] and [Dy/Fe]. This data may help to constrain chemical evolution 
models of Milky Way \emph{r-} and \emph{s-}process elements and the yields of 
massive stars. We emphasize that including yields of neutron-star or black hole 
mergers is now crucial if we want to quantitatively compare observations to 
Galactic chemical evolution models.}


\maketitle

\section{Introduction}

The surface abundances of FGK dwarf stars do not show major 
changes along their main sequence evolution, reflecting their original pristine 
chemical composition. The Milky Way stars observable today were created thanks 
to a gas which had been enriched by several generations of stars, or from  the 
in-fall or accretion of gas in the Galaxy. On the one hand, many comprehensive 
studies were able to constrain the chemical evolution of $\alpha$ and iron peak 
elements in the Milky Way, thanks to recent massive spectroscopic surveys like 
RAVE \citep{steinmetz_2003}, APOGEE \citep{wilson_2010}, and Gaia-ESO Survey 
\citep{gilmore_2012}, but also thanks to more classical studies, for example
\citet{Adibekyan2011} and \citet{Haywood_2013}. On the other hand, less theoretical and 
observational data are available for elements heavier than $Z\sim35$, usually 
called neutron-capture elements because they are formed by the addition of neutrons in 
stellar interiors. To create such nuclei, two main processes are known, first through the slow 
neutron-capture (\emph{s-}process) or rapid neutron-capture (\emph{r-}process), 
depending on whether the neutron-capture is slow or rapid with respect to the timescale 
of the $\beta$ decay \citep{burbidge_1957}.

The \emph{s-}process  known to take place in the He-burning core of massive stars 
and in the convective C-burning shell \citep{peters_1968, lamb_1977} is also called the 
weak \emph{s-}process. The \emph{s-}process also happens in the asymptotic giant branch (AGB) 
of lower mass stars ($M<4M_{\sun}$) at solar and lower metallicities \citep{bisterzo_2011}, 
also denoted the main \emph{s-}process. By ejecting their envelope, AGB stars are thought 
to be the main contributors for enriching the interstellar medium (ISM) in nuclei 
with atomic mass number $90<A<204$. Finally, the last \emph{s-}process, also called the 
strong \emph{s-}process, is responsible for half of the solar ${}^{208}\text{Pb}$ and 
takes place in low-metallicity AGB stars \citep{travaglio_2001_AGB_s}. In this study, 
as detailed later, we  focus on barium, which is a heavy \emph{s-}process element, 
of the second peak, mostly synthesized via the main \emph{s-}process.

The rapid neutron 
capture process takes place on a much shorter timescale with respect to the $\beta$ 
decay and when the density of neutrons is high enough. The \emph{r-}process elements are 
basically divided into three peaks: A=80, 130, and 194, depending on the timescale of the 
neutron flow and the atomic structure \citep{seeger_1965}.
At least two production sites involving core-collapse  supernovae (CCSN) have been proposed for \emph{r-}process elements: 
neutrino induced winds via the so-called weak \emph{r-}process, 
but they present limitations in producing nuclei with atomic number $A>100$ \citep{woosley_1994} and 
magneto-hydrodynamic jet explosions via the so-called main \emph{r-}process 
(for $A>130$).  The main \emph{r-}process
  contributes to the ISM enrichment with 
significant yields of $10^{-3}-10^{-2}M_{\sun}$ in \emph{r-}process material for typical 
initial mass of $13 \lesssim M \lesssim 25\,M_{\sun}$ \citep{nishimura_2006, nishimura_2015} 
over a typical timescale of few hundred million years. Unfortunately, such yields suffer 
from larger uncertainty, and are very mass and metallicity dependent. The main \emph{r-}process 
is also supposed to occur during the  merging of black holes or neutrons stars (NS, 
\citealt{freiburghaus_1999}). This has been confirmed observationally by the recent LIGO/Virgo 
detection of the first gravitational wave signal produced by NS-NS merging 
\citep{abbott_kilo_2017, abbott_2017b}; however, no robust yields are available for this 
mechanism.

The chemical evolution of such heavy elements in the Milky Way is still then a matter of 
debate. Strong efforts on the theoretical side have been made in order to trace the origin 
of heavy elements in the Galaxy. For example, \citet{travaglio_1999} investigated the evolution 
of heavy elements from Ba to Eu in the thin disc, thick disc, and halo. \citet{bisterzo_2017} 
followed the same approach, but focused on the so-called ${}^{13}\text{C}$-pocket, a major 
source of uncertainty in ABG yields. In this study our goal is to understand the chemical 
evolution of three almost pure \emph{r-}process elements: europium (Eu), gadolinium (Gd), 
and dysprosium (Dy). Europium can be considered  a pure \emph{r-}process element because 98\% 
of the solar europium comes from the \emph{r-}process \citep{sneden_2008}, and  gadolinium and 
dysprosium as well, with 82\% and 88\% of the solar abundance coming from the \emph{r-}process. 
Moreover, a powerful element that can be used to provide more constraints on the chemical 
evolution of such pure \emph{r-}process nuclei is  barium (Ba). Barium is for the most part an 
\emph{s-}process element as 84\% of the solar Ba originates from the \emph{s-}process 
\citep{sneden_2008}. The [element/Ba] abundance ratios provide  a direct way to quantify 
the relative importance of the \emph{r-} and \emph{s-} channels during the evolution of the 
Galaxy. Eu, Gd, and Dy are basically supposed to be produced via the same channel because of 
their very similar atomic mass number, and are located between the second and third peaks of 
the \emph{r-}process. So we do not expect any major differences in the chemical evolution of these 
almost pure \emph{r-}process elements. Because the unclear physical conditions of the 
astrophysical sites lead to very uncertain \emph{r-}process predictions, Galactic chemical 
evolution models of \emph{r-}process elements are quite challenging. Also, in up-to-date models, 
for elements heavier than Ba, the solar \emph{r-}process contribution is directly deduced from 
the \emph{s-} fraction of the solar abundances, giving some limitations to the GCE model 
\citep{kappeler_2011, prantzos_2018}. Additionally, rotation in massive stars is now known to 
play a key role in the efficiency of the stellar yields \citep{chiappini_2006, prantzos_2018}. 

Ba and Eu have been observationally studied  in the main Milky Way components. For 
example, \citet{battistini_bensby_2016} recently studied the temporal evolution of such 
elements in the thin and the thick discs for 400 stars, while \citet{delgado_mena_2017} 
presented Ba and Eu abundances for $\sim6\,00$ FGK stars, and also studied the halo.

On the contrary, Gd and Dy have been poorly studied in the the Milky Way disc. 
\citet{overbeek_2016} measured Gd and Dy for 68 stars in 23 open clusters while \citet{spina_2018} 
studied the temporal evolution of Gd and Dy for 79 solar twin stars.

We aim here to study homogeneously the evolution of Ba, Eu, Gd, and Dy for a statistically 
significant sample of stars, covering a large domain of metallicity. To this end we 
automatically derived a very large set of  abundances for these elements, thanks to ESO archive spectra, 
for a sample of $\sim6\,000$ stars. This study is placed in the context of the AMBRE Project 
\citep{delaverny_2013}. In order to put more constraints on the \emph{r-}process elements 
origin, we also focus our study on the two main components of the Milky Way: the thin and 
thick discs.

The paper is organized as follows. In Sect.~\ref{observationnal_data} we present the 
spectroscopic data used for our analysis, while in Sect.~\ref{metohd} we detail our 
automatic procedure of abundance determination. In Sect.~\ref{ambre_catalogue}, we 
validate and present our AMBRE catalogue of Ba, Eu, Gd, and Dy abundances, while in 
Sect.~\ref{working_sample} we define our working stellar sample. The chemical evolution 
of Ba, Eu, Gd, and Dy in the Milky Way disc is discussed in Sect.~\ref{discussion} in the 
context of recent chemical evolution models. We conclude this work in 
Sect.~\ref{conclusiooooonnnn}.

\section{Observational data set from the AMBRE Project}\label{observationnal_data}

This work is based on ESO archived spectra data from the AMBRE project. We recall 
that this project is dedicated to the parametrization of the HARPS, FEROS, UVES, and 
GIRAFFE spectral archives \citep{delaverny_2013}, providing robust automatic determinations 
of the radial velocity ($\vrad$), effective temperature ($\tef$), surface gravity ($\logg$), 
metallicity ($\mh$), and global $\alpha$ enrichment with respect to iron ($\alffe$) together 
with their associated errors. The present study focuses on a subsample of the first three 
spectral sets that have already been parametrized: HARPS \citep{depascale_2014}, FEROS 
\citep{worley_2012}, and UVES \citep{worley_2016}.

These analysed subsamples consist of spectra with a good AMBRE quality flag (lower or equal 
to 1; see e.g. \citealt{worley_2012} for details on this label). The typical total errors on 
$\tef$, $\logg,$ and $\mh$ are [$108\,$K, $0.16\,$\cms, $0.10\,$dex] for UVES, [$93\,$K, 
$0.26\,$\cms, $0.08\,$dex] for HARPS, and [$120\,$K, $0.20\,$\cms, $0.10\,$dex] for FEROS. 
In the following, we  also use the AMBRE estimates of the signal-to-noise ratio ($\snr$) 
and the FWHM of the cross-correlation function estimated when deriving $\vrad$ for a given 
star ($\ccf$).

\section{Automatic abundance analysis of n-capture elements}\label{metohd}

The \emph{r-} and \emph{s-}abundances of the AMBRE spectra were automatically derived 
via an optimization method by coupling a pre-computed synthetic spectra grid and the 
GAUGUIN Gauss--Newton algorithm. This method is presented in detail in \citet{guiglion_2016}, 
but we give here a brief summary of the procedure, focusing mainly on the line-list adopted 
for the derivation of $\ion{Ba}{II}$, $\ion{Eu}{II}$, $\ion{Gd}{II}$, and 
$\ion{Dy}{II}$\footnote{In order not to crowd the text and figures, we  adopt the notation 
`Ba' for $\ion{Ba}{II}$ (also for the other elements).}. 

The main idea was to identify reliable atomic \emph{r-} and \emph{s-}process lines 
common to the FEROS ($3\,500-9\,200\,$\AA), HARPS ($3\,780-6\,910\,$\AA), and UVES spectral 
domains. For UVES, we took advantage of three different set-ups: Red580 ($4\,726-6\,835\,$\AA), 
U564 ($4\,583-6\,686\,$\AA) and U437 ($3\,731-4\,999\,$\AA). The adopted lines and their 
spectral synthesis properties are as follows (see also \tablename~\ref{table_localisation_line}):\\

- Barium suffers from strong hyperfine splitting, and we adopted the lines and the hyperfine and 
isotopic structure from \citet{rutten_1978} for the lines $5\,853.69$, $6\,141.73$, 
$6\,496.90\,$\AA, including the following isotopes: $\text{Ba}^{130,~132,~134,~135,~136,~137,~138}$. 
We note that these isotopes are included in the spectral synthesis (with solar isotopic ratios), but 
we do not derive individual isotopic abundances (because of too low spectral resolution and high enough signal-to-noise ratio).\\

- Europium has two main isotopes, $\text{Eu}^{151}$ and $\text{Eu}^{153}$ and suffers from 
strong hyperfine splitting. We adopted the hyperfine and isotopic structure from the line-list 
of \citet{lawler_2001_eu} for the four spectral lines considered: $4\,129.72$, $4\,205.04$, 
$6\,437.64$, and $6\,645.13\,$\AA.\\

- Gadolinium shows weaker lines compared with europium, and uncertain hyperfine splitting 
data can be found in the literature. We therefore do not take into account such structures. 
We adopted five lines from the work of \citet{den_hartog_2006}: $4\,085.57$, $4\,191.08$, 
$4\,316.08$, $4\,483.32$, and $4\,498.28\,$\AA.\\

- Weak contribution of the hyperfine splitting is predicted for dysprosium so we do not take 
it into account in our spectral synthesis. We adopted the atomic data from \citet{wickliffe_2000} 
for the two lines considered: $4\,073.12$, $4\,449.70$.\\

We note that all lines are not observed, detected, and unblended in each star.

\begin{table}
\centering
\begin{tabular}[c]{c c c c c c c}
El. & line (\AA) & $\loggf$ & $\chi_e$ & HFS & Ref. & Spec. \\
\hline
\hline
$\ion{Ba}{II}$ & 5853.69 & -1.01 & 0.60 & Yes & $ru$ & H/F/U564 \\
$\ion{Ba}{II}$ & 6141.73 & -0.07 & 0.70 & Yes & $ru$ & H/F/U564 \\
$\ion{Ba}{II}$ & 6496.90 & -0.38 & 0.60 & Yes & $ru$ & H/F/U564 \\
$\ion{Eu}{II}$ & 4129.72 & +0.22 & 0.00 & Yes & $la$ & H/F/U437 \\
$\ion{Eu}{II}$ & 4205.04 & +0.21 & 0.00 & Yes & $la$ & H/F/U437 \\
$\ion{Eu}{II}$ & 6437.64 & -0.32 & 1.31 & Yes & $la$ & H/F/U580 \\
$\ion{Eu}{II}$ & 6645.13 & +0.12 & 1.38 & Yes & $la$ & H/F/U580 \\
$\ion{Gd}{II}$ & 4085.56 & -0.01 & 0.73 & No  & $dh$ & H/F/U437 \\
$\ion{Gd}{II}$ & 4191.05 & -0.48 & 0.43 & No  & $dh$ & H/F/U437 \\
$\ion{Gd}{II}$ & 4316.05 & -0.45 & 0.66 & No  & $dh$ & H/F/U437 \\
$\ion{Gd}{II}$ & 4483.32 & -0.42 & 1.06 & No  & $dh$ & H/F/U437 \\
$\ion{Gd}{II}$ & 4498.28 & -1.08 & 0.43 & No  & $dh$ & H/F/U437 \\
$\ion{Dy}{II}$ & 4073.12 & -0.32 & 0.54 & No  & $wl$ & H/F/U437 \\
$\ion{Dy}{II}$ & 4449.70 & -1.03 & 0.00 & No  & $wl$ & H/F/U437 \\
\hline
\end{tabular}
\caption{\label{table_localisation_line}Element, wavelength, $\loggf$, hyperfine 
structure, and reference for the 15 spectral lines used in this study. $ru$ == 
\citet{rutten_1978}. $la$ == \citet{lawler_2001_eu}. $dh$ == \citet{den_hartog_2006}. 
$wl$ == \citet{wickliffe_2000}. For a given line the available spectrograph is 
indicated (Spec): H, F, and U stand for HARPS, FEROS, and  UVES(+set-up), respectively.}
\end{table}

We then adopted a line-list for the atomic blends from the Vienna Atomic Line 
Database (VALD3, \citealt{kupka_1999, kupka_2000}) over the domains covered by 
the adopted lines for the abundance analysis. Additionally, the line-lists of 
twelve molecular species were also taken into account: CN \citep{Sneden2014}, 
TiO (Plez, priv. comm.), C2 \citep{Brooke2013, Ram2014}, CH \citep{2014A&A...571A..47M}, 
ZrO (Plez, priv. comm.), OH and NH (Masseron, priv. comm), CaH (Plez, priv. comm.), 
VO (Plez, priv. comm.), FeH \citep{dulick_2003}, MgH \citep{hinkle_2013}, and SiH 
\citep{1992RMxAA..23...45K}. \\

Based on these line lists, a specific synthetic spectra grid was computed using 
the MARCS model atmosphere \citep{gustafsson_2008} and the LTE TURBOSPECTRUM code 
\citep{plez_2012}. Five dimensions were considered for this grid: $\tef$, $\logg$, 
$\mh$, $\alffe,$ and $[\text{X}/\text{Fe}]$, where $[\text{X}/\text{Fe]}$ correspond 
the \emph{r-} and \emph{s-} enhancements. The ranges of the atmospheric parameters are 
those of the AMBRE grid \citep{laverny_2012}, $4\,000\le\tef\le8\,000\,$K (in steps of 
$250\,$K), $+0\le\logg\le+5.5\,$\cms\ (in steps of $0.5\,$\cms), $-5\le\mh\le+1\,$dex, 
whereas the enhancement in \emph{r-} and \emph{s-} varies over a range of $\pm1.2\,$dex 
around the metallicity in steps of $+0.2\,$dex (13 different values of $[\text{X}/\text{Fe}]$).

For the present grid, a specific microturbulence velocity law (polynomial variation as 
a function of $\tef$, $\logg,$ and $\feh$) has been adopted for the synthetic spectrum 
calculation, as was done in our computation of Gaia ESO Survey grids. Moreover, consistent 
$\alffe$ enrichments for the model atmosphere and the synthetic spectrum calculations 
were considered.

The micro-turbulence velocity ($\xi$) was included in the grid computation by varying 
$\xi$ as a function of $\tef$, $\logg,$ and $\feh$ as adopted in the Gaia-ESO Survey 
(Bergemann et al., in preparation; based on $\xi$ determinations from literature samples). 
The total number of synthetic spectra is $174\,534$, computed on a wavelength range of 
$40\,$\AA, centred on each of the spectral lines presented in \tablename~\ref{table_localisation_line}, 
adopting a sampling of $0.004\,$\AA.

We then interpolate the pre-computed 5-D synthetic spectra grid presented above at 
the atmospheric parameters of the targets ($\tef^{\star}$, $\logg^{\star}$, $\feh^{\star}$, 
and $\alffe^{\star}$) derived within the AMBRE Project to prepare a small set of interpolated 
synthetic spectra for a direct comparison with the observation. The resulting 1-D grid in 
abundance at $\tef^{\star}$, $\logg^{\star}$, $\feh^{\star}$, and $\alffe^{\star}$ varies 
from $-1.2$ to $+1.2\,$dex around the metallicity of the star and is composed of 13 model 
spectra. The resolution of the observed spectra was degraded to $40\,000$ for UVES and FEROS, 
while for HARPS, we kept the original spectral resolution of $R=110\,000$, re-sampling these 
spectra to a pixel size of $0.05\,$\AA~and $0.015\,$\AA, respectively. The same convolution 
and re-sampling was performed for the synthetic spectra grid in order to perform the abundance 
determination by automatically comparing the observed and synthetic spectra. Finally, an 
automatic adjustment of the continuum and the correction of the radial velocity was performed 
on the observed spectrum as already described in \citet{guiglion_2016}. 

For a given spectrum, from the 1-D grid described above, we compute a quadratic distance 
between the observed spectrum and each point of the 1-D grid. For each spectral line 
presented in \tablename~\ref{table_localisation_line}, we focused on a small wavelength 
range around the line, basically $\pm0.2/0.3\,$\AA. The minimum of distance provides a 
first estimate of the solution, then this first guess is optimized via the Gauss--Newton 
algorithm GAUGUIN (see \citealt{guiglion_2016} for more details). Upper limits are provided 
when the spectral feature is too weak with respect to the $\snr$ of the spectrum. Finally, 
we assume for this abundance analysis that all the targets are single stars since binary 
detection is not a part of the AMBRE parametrization pipeline. However, we  point out that 
most of the spectroscopic binaries present in the original sample should have been rejected 
when selecting only good parametrized spectra.

\subsection{Solar scale and average abundance calculation}
No \emph{s-} and \emph{r-}process abundances are available for the benchmark stars in 
the literature. To calibrate our abundances, we searched for solar spectra in the FEROS, 
HARPS, and UVES samples. We computed an average solar abundance for the available solar 
spectra (43 for HARPS and FEROS, and 6 for UVES) for each spectral line of 
\tablename~\ref{table_localisation_line} for each spectrograph. We discarded measurement presenting 
too low S/N, limits, and too large errors, leading to 22 spectra. Basically, differences 
(biases) with respect to the solar values \citep{grevesse_2007} are of the order of 0.15 
dex for Ba, 0.10 dex for Eu, and 0.25 dex for Gd and Dy, in absolute value, and vary from one 
spectrograph to the other. The typical dispersion is quite weak, around 0.12 for Ba corrections and 
0.10 dex for Eu, Gd, and Dy corrections. Such biases results from mismatches 
between solar synthetic and observed spectra due probably to the uncertainties in the adopted 
line data. We recall that no astrophysical calibration of our line-list has been preformed 
since we favored our {a-posteriory} calibration of the abundances. We note that a similar 
approach has been adopted by the Gaia-ESO Consortium.

Then, for each stellar sample, we computed for each chemical element an average of the 
available lines (only true measurements, no upper limits) and they have been put on the 
solar scale thanks to the biases mentioned above.

\subsection{Error budget}\label{error_budget}
In order to derive proper uncertainties on the Ba, Eu, Gd, and Dy abundances, we combined 
two main sources of uncertainty: propagation of the errors of the atmospheric parameters and 
line-to-line scatter for a given element. We first propagated the errors on the three 
atmospheric parameters $\big\{\tef^{\star},\,\logg^{\star},\,\mh^{\star}\big\}$ provided by 
AMBRE and summed them quadratically, leading to a first error term $\text{e}_{[\text{X/Fe}]}$. 
For a given element with several lines abundance measured, we also computed their standard 
deviation. This leads to a second error source, denoted $\sigma_{[\text{X/Fe}]}$. Quadratically 
summing $\text{e}_{[\text{X/Fe}]}$ and $\sigma_{[\text{X/Fe}]}$ gives us the final uncertainty, 
denoted $\text{e}_{\text{tot}}[\text{X/Fe}]$ (which is probably overestimated). By applying cuts 
on these errors,  in the next section we  present our working samples for our science application.

We also estimated the impact of a bad continuum placement. To do so, we modified the continuum 
of a synthetic solar spectrum of about $3\%$, thanks to a third-order polynomial function
\footnote{We also tested   a first- and second-order polynomial function, and the 
results are roughly identical.}, around the lines used for the chemical abundance analysis. These 
tests were done at $\snr=100$, for $1\,000$ noisy realizations, at both $R=40\,000$ (UVES/FEROS 
like resolution) and $R=100\,000$ (HARPS-like resolution). The typical errors induced by the bad 
continuum placement are of the order of 0.04 dex for Ba, 0.08 dex for Eu, and 0.12 dex for Gd and Dy. 
These tests are very pessimistic  because in practice our automatic procedure is able to 
renormalize to a precision better than $1\%$ of the continuum for $\snr>15$. The resulting errors will then be 
negligible with respect to those due to the atmospheric parameters and line-to-line scatter.

\subsection{Repeated observations}
As presented in the next section, the FEROS sample is composed of 5981 spectra, including 
repeated observations. In order to show the robustness of the AMBRE \emph{r-} and \emph{s-} process abundances, 
we present average abundances and typical dispersion for some dwarfs/subgiants with repeated 
observations ($\text{N}_\text{{rep}}>10$) in \tablename~\ref{table_reape}. We first note from 
this table that the dispersions on the atmospheric parameters are much smaller than the errors on 
these quantities estimated during the AMBRE parametrization. This confirms that our first error 
term ($\text{e}_{[\text{X/Fe}]}$) is overestimated and it refers mainly to the external error, 
not to the internal relative error. Then, the typical dispersion of the abundances is around or 
well below 0.10 dex, for the four elements, even in the metal-poor regime, for example HD203608. 
We note that this dispersion can be  explained by the fact that for a given star, all the 
repeats do not have the same atmospheric parameters. In general, Gd and Dy show higher dispersion 
principally because of weaker spectral features. We observe the same trends for repeated observations 
in the HARPS sample, and the samples UVES580 and UVES437 for Eu, Gd, and Dy, and in the sample 
of UVES564 with Ba abundances. 

\begin{sidewaystable}
\centering
\begin{tabular}[c]{c c c c c c c c c c c c c c}
Star & $\text{N}_\text{{rep}}$ & $\langle\tef\rangle \pm \sigma$ & $\langle\logg\rangle \pm \sigma$ & $\langle\mh\rangle \pm \sigma$ & $\langle\text{[Ba/Fe]}\rangle \pm \sigma \pm \langle \text{e}_{\text{tot}}\rangle$ & $\langle\text{[Eu/Fe]}\rangle \pm \sigma \pm \langle \text{e}_{\text{tot}} \rangle$ & $\langle\text{[Gd/Fe]}\rangle \pm \sigma \pm \langle \text{e}_{\text{tot}} \rangle$ & $\langle\text{[Dy/Fe]}\rangle \pm \sigma \pm \langle \text{e}_{\text{tot}} \rangle$ \\
     &   & (K) & \cms & dex & dex & dex & dex & dex \\
\hline
\hline
HD47875   &  13 & $5804 \pm 34$ & $4.54 \pm 0.02$ & $+0.05 \pm 0.04$ & $+0.13 \pm 0.08 \pm 0.19$ & $-0.01 \pm 0.08 \pm 0.23$ & $+0.18 \pm 0.11 \pm 0.43$ & $-0.21 \pm 0.15 \pm 0.07$ \\
HD70573   &  24 & $5937 \pm 19$ & $4.53 \pm 0.05$ & $+0.02 \pm 0.02$ & $-0.26 \pm 0.03 \pm 0.18$ & $-0.04 \pm 0.08 \pm 0.25$ & $+0.32 \pm 0.10 \pm 0.42$ & $-0.27 \pm 0.11 \pm 0.03$ \\
HD4128    &  28 & $5157 \pm 19$ & $3.12 \pm 0.03$ & $+0.14 \pm 0.02$ & $+0.72 \pm 0.01 \pm 0.08$ & $+0.09 \pm 0.03 \pm 0.17$ & $-0.32 \pm 0.07 \pm 0.08$ & $-0.02 \pm 0.03 \pm 0.31$ \\
HD102870  &  23 & $6111 \pm  5$ & $4.18 \pm 0.01$ & $+0.16 \pm 0.01$ & $-0.07 \pm 0.03 \pm 0.11$ & $-0.05 \pm 0.10 \pm 0.11$ & $-0.19 \pm 0.24 \pm 0.34$ & $-0.24 \pm 0.20 \pm 0.02$ \\
HD96064   &  17 & $5569 \pm  9$ & $4.67 \pm 0.01$ & $+0.10 \pm 0.01$ & $+0.14 \pm 0.02 \pm 0.08$ & $+0.07 \pm 0.05 \pm 0.15$ & $+0.33 \pm 0.08 \pm 0.17$ & $-0.36 \pm 0.07 \pm 0.08$ \\
HD203608  &  64 & $6005 \pm  6$ & $4.07 \pm 0.01$ & $-0.77 \pm 0.01$ & $-0.11 \pm 0.01 \pm 0.11$ & $+0.23 \pm 0.07 \pm 0.24$ & $+0.25 \pm 0.11 \pm 0.04$ & $+0.08 \pm 0.10 \pm 0.03$ \\
HD15526   &  12 & $5730 \pm 66$ & $4.71 \pm 0.08$ & $+0.01 \pm 0.03$ & $+0.15 \pm 0.09 \pm 0.09$ & $+0.16 \pm 0.13 \pm 0.29$ & $+0.50 \pm 0.06 \pm 0.14$ & $-0.21 \pm 0.04 \pm 0.12$ \\
Gl667     &  29 & $4654 \pm  8$ & $4.62 \pm 0.02$ & $-0.46 \pm 0.01$ & $-0.19 \pm 0.03 \pm 0.03$ & $+0.27 \pm 0.05 \pm 0.25$ & $+0.42 \pm 0.14 \pm 0.62$ & $+0.01 \pm 0.10 \pm 0.27$ \\
HD212301  &  32 & $6172 \pm 20$ & $4.37 \pm 0.01$ & $+0.12 \pm 0.02$ & $-0.22 \pm 0.04 \pm 0.13$ & $+0.01 \pm 0.07 \pm 0.18$ & $-0.03 \pm 0.16 \pm 0.33$ & $-0.29 \pm 0.12 \pm 0.04$ \\
HD75289   &  46 & $6116 \pm  6$ & $4.29 \pm 0.01$ & $+0.29 \pm 0.01$ & $-0.09 \pm 0.01 \pm 0.09$ & $-0.01 \pm 0.02 \pm 0.08$ & $-0.08 \pm 0.05 \pm 0.42$ & $-0.23 \pm 0.05 \pm 0.06$ \\
HD217107  & 103 & $5624 \pm  6$ & $4.44 \pm 0.01$ & $+0.26 \pm 0.01$ & $-0.12 \pm 0.01 \pm 0.02$ & $+0.11 \pm 0.03 \pm 0.18$ & $-0.25 \pm 0.07 \pm 0.17$ & $-0.43 \pm 0.06 \pm 0.08$ \\
\hline
\end{tabular}
\caption{\label{table_reape}Mean atmospheric parameters, and Ba, Eu, Gd, and Dy abundances 
for some examples of FEROS stars with repeated observations. The dispersions ($\sigma$) 
over the $\text{N}_\text{{rep}}$ are presented, as well as the mean total error 
$\langle \text{e}_{\text{tot}} \rangle$ for the abundances.}
\end{sidewaystable}

When cross-matching FEROS and HARPS together, 34 subgiants and dwarf stars share 
spectra with both spectrographs, covering a metallicity range from -0.47 to $+0.20\,$dex. 
For 19 of them we are able to measure abundances. Basically, for a given star, the dispersion 
between the repeats is again below $0.1\,$dex for Eu and Ba abundances, and of the order 
of $0.10/0.15\,$dex for Gd and Dy, showing good reliability of the adopted AMBRE atmospheric 
parameters and abundances derived. For the remaining 15 stars, only upper limits are available, 
but are fully consistent between FEROS and HARPS spectra for a given star.

\section{AMBRE catalogue of Ba and \emph{r-}process abundances}\label{ambre_catalogue}

In this section we present how we combined the three different samples of HARPS, UVES, and 
FEROS in order to provide a catalogue of Ba and \emph{r-}process abundances. We also 
present abundances for 19 identified Gaia benchmark stars in our catalogue. We finally 
confirm the reliability of our measurements by comparing our abundances with external data sets.

\subsection{Combining UVES, FEROS, and HARPS samples}

We now have in hand three different samples, HARPS, FEROS, and UVES, containing a few 
repeated observations. 

For FEROS, we performed a cross-match on the spectra coordinates with a radius of $10\,$arcsec 
on the sky, leading to a remaining sample composed of $3\,526$ stars. For a given star with several 
spectra collected with the same spectrograph, we computed averaged atmospheric parameters and 
averaged chemical abundances, leading to a better precision. For HARPS, we recall that we adopted 
the sample of $4\,355$ stars from \citet{mikolaitis_2017} based on a search of  coordinates 
and of atmospheric parameter differences, containing then individual stars. 

For HARPS and FEROS data, it was easy to combine the two samples because Ba, Eu, Gd, and Dy were 
measured homogeneously with the same spectral lines. In order to eliminate repeats between the two 
samples, we made a second cross-match with a radius of $10\,$arcsec on the sky over the coordinates 
of the two samples, leading to a remaining sample of 5808 stars (2281 HARPS stars and 3527 
FEROS stars). Here again, for a star with several repeats, we computed a mean of its atmospheric parameters 
and chemical abundances, derived with all the available HARPS and FEROS spectra. 

As presented in \tablename~\ref{nb_star_elem_spectro}, in UVES, Ba was derived thanks to the 
set-up U564 for 528 spectra, Eu thanks to U437 (1414 spectra) and U580 (3628 spectra), Gd thanks 
to U437, and Dy thanks to U437 and U580. For each set-up, we searched for repeated observations and 
we performed a cross-match with a radius of $2\,$arcsec on the sky, resulting in 258 individual 
stars in U654, 744 in U437, and 1030 in U580. We note  that there are 213 common stars between 
U437 and U580, showing consistent atmospheric parameters between the two set-ups, with a typical 
dispersion of $81\,$K in $\tef$, $0.21\,$\cms~in $\logg$ and $0.09\,$dex in $\mh$. The Eu abundances 
are also consistent within a 1$\sigma$ error, beside the fact that different Eu spectral lines are 
used in each set-up.

As we assume that the stars are slow rotators, and do not include rotation in our procedure based 
on a synthetic spectra grid, we exclude stars with high FWHM of the cross-correlation function ($\ccf$) 
computed during the AMBRE parametrization. The $\ccf$ gives  partial information on the 
rotational velocity. We followed the same criteria as in \citet{guiglion_2016}, excluding stars with 
$\ccf>20\,$\kms~for FEROS targets, and $\ccf>15\,$\kms~for HARPS and UVES targets. We recall that in their 
study \citet{guiglion_2016} established these criteria estimating the $\vsini$ effect on the lithium 
abundance measurement, spanning a wide range of line strengths and depths. In our study, we span a wide 
range of spectral line profiles, from strong lines for Ba and Eu to weak lines for Dy and Gd, allowing 
us to apply these criteria. These high rotation rate stars  are in the minority (around 5\%), so the cuts made 
here do not affect the global statistics that much. We also note that we excluded stars with abundance 
uncertainty $\text{e}_{\text{tot}}$ higher than 1 dex, removing then 5\% of the sample. Finally, our 
pipeline did not converge for  $\sim5\%$ of the sample.

Finally, the number of stars per chemical element are presented in \tablename~\ref{nb_star_elem_spectro}. 
The mean $\snr$ is 130 for UVES and FEROS, and 50 for HARPS. Our AMBRE Ba and \emph{r-}process catalogue 
contains both dwarf and giant stars. It is the first time that such a catalogue has been created, coupling a high 
statistics, and a wide coverage in atmospheric parameters. The AMBRE catalogue of Ba, Eu, Gd, and Dy 
abundances is presented in \tablename~\ref{table_ambre_ba}. In Sect.~\ref{working_sample} 
we build a working subsample from our catalogue.

\begin{sidewaystable}
\centering
\begin{tabular}[c]{c c c c c c c c c c}
Identifier & $\tef$ & $\logg$ & $\mh$ & $\alffe$ & $\snr$ & [Ba/Fe] & $\text{e}_{\text{tot}}$ & $N_{line}$ & spectro.\\
           &  K     & \cms   & dex   & dex &     &  dex    & dex &  &  \\
\hline
\hline
HD125276 & 5946 & 3.99 & -0.77 & +0.12 & 211 & +0.17 & 0.21 & 3 & HARPS \\
HD150177 & 6005 & 3.48 & -0.84 & +0.17 & 191 & +0.09 & 0.12 & 2 & HARPS \\
HD693    & 6031 & 3.70 & -0.37 & +0.12 & 173 & -0.10 & 0.14 & 3 & HARPS \\
HD22879  & 5693 & 3.82 & -1.01 & +0.33 & 168 & +0.05 & 0.08 & 1 & HARPS \\
HD25704A & 5804 & 3.82 & -1.00 & +0.26 & 62  & <+0.66 & - & - & HARPS \\
HD8558   & 5742 & 4.80 & -0.12 & -0.04 & 61  & <-0.22 & - & - & HARPS \\
..   & .. & .. & .. & .. & ..  & .. & .. & .. & .. \\
\hline
\hline
\end{tabular}
\caption{\label{table_ambre_ba}Identifier, atmospheric parameters, $\snr$, Ba abundance,  
$\text{e}_{\text{tot}}$, number of Ba line used, and spectrograph. When no error and number of 
lines are indicated, the abundance is an upper limit. The full abundance tables 
of Ba, Eu, Gd, and Dy are available at the CDS.}
\end{sidewaystable}

\begin{table}
\centering
\begin{tabular}[c]{c c c c c c c}
      & HARPS & FEROS & U564 & U437 & U580 & $\sum$ \\
\hline
\hline
Ba   &  $1\,911$  & $2\,951$  & $195$  &   -   &  -     & $5\,057$ \\
Eu   &  $1\,880$  & $3\,104$  &   -    & $363$ &  $921$ & $6\,268$ \\
Gd   &  $1\,946$  & $3\,108$  &   -   &  $377$ &    -   & $5\,431$ \\
Dy   &  $2\,015$  & $3\,091$  &   -   &  $373$ &    -   & $5\,479$ \\
\hline
\end{tabular}
\caption{\label{nb_star_elem_spectro}Number of stars for each derived abundance 
(including upper limits) and spectrograph.}
\end{table}

\subsection{Ba, Eu, Gd, and  Dy of the Gaia-benchmark stars}\label{gaia_bench_abund}

In our samples, we identified several Gaia-benchmark stars \citep{jofre_2013}. This identification 
was performed with the coordinates and TARGNAME identifier, resulting in 
19 stars, for which we were able to derive Ba, Eu, Gd, and Dy (including upper limits) 
using the AMBRE atmospheric parameters.
We present our results in \tablename~\ref{table_benchmarks}. It is the first time that such a table of 
\emph{s-} and \emph{r-}process abundances in the Gaia-benchmark 
stars has been published. We note that the uncertainty goes typically from 0.1 to 0.3 dex, but can suffer 
from larger errors, for example Dy in $\tau$ Cet ($[\text{Dy}/\text{Fe}]=+0.22\pm0.45\,$dex). The main 
reason is that for lower metallicity stars, spectral lines  start to be too weak to be accurately measured 
and/or fewer lines are available for the analysis. The same is also true for too hot stars. We finally 
note that Eu and Gd Arcturus abundances are fully consistent within $1\sigma$ with \citet{overbeek_2016}. Eu 
is also in a very good agreement within $1\sigma$ with \citet{vanderswaelmen_2013} and \citet{worley_2009}.

\begin{table*}
\centering
\begin{tabular}[c]{c|c c c c c c c c c c c c}
Star           & $\tef$   &$\logg$& $\mh$ & [Ba/Fe] & $N_{\text{l}}$ & [Eu/Fe] & $N_{\text{l}}$ & [Gd/Fe] & $N_{\text{l}}$ & [Dy/Fe]& $N_{\text{l}}$ & Spec. \\
\hline
\hline
\bf{F Dwarf}   &          &      &       &                   &   &                   &   &                   &   &                   &   &   \\
Procyon        & $6\,424$ & 3.81 & -0.34 & $+0.30_{\pm0.25}$ & 3 & $+0.20_{\pm0.11}$ & 3 & $+0.11_{\pm0.47}$ & 4 & $+0.13_{\pm0.28}$ & 2 & U \\
HD 49933       & $6\,482$ & 3.90 & -0.54 & $-0.40_{\pm0.37}$ & 2 & $-0.07_{\pm0.49}$ & 3 & $+0.67_{\pm0.50}$ & 3 & $+0.18_{\pm0.45}$ & 2 & H \\
HD 84937       & $6\,300$ & 3.71 & -2.33 & $-$               & 0 & $<+1.70$          & 1 & $<+1.70$          & 2 & $<+1.30$          & 1 & U \\
\hline 
\bf{FGK Subgiants}&          &      &       &                &   &                   &   &                   &   &                   &   &   \\
$\delta$ Eri   & $5\,033$ & 3.82 & +0.09 & $-0.11_{\pm0.15}$ & 3 & $+0.02_{\pm0.07}$ & 4 & $-0.05_{\pm0.32}$ & 5 & $-0.04_{\pm0.45}$ & 2 & H \\
HD 140283      & $5\,700$ & 3.48 & -2.52 & $<+2.00$          & 2 & $<+1.60$          & 1 & $<+1.60$          & 1 & $<+1.40$          & 1 & F \\
$\epsilon$ For & $5\,041$ & 3.42 & -0.67 & $+0.04_{\pm0.33}$ & 3 & $+0.37_{\pm0.10}$ & 3 & $+0.49_{\pm0.22}$ & 3 & $+0.47_{\pm0.35}$ & 2 & F \\
$\beta$ Hyi    & $5\,775$ & 4.02 & -0.11 & $+0.17_{\pm0.21}$ & 3 & $+0.05_{\pm0.24}$ & 2 & $-0.05_{\pm0.25}$ & 3 & $+0.15_{\pm0.37}$ & 2 & F \\
\hline
\bf{Solar-type}&          &      &       &                   &   &                   &   &                   &   &                   &   &   \\
$\alpha$ Cen A & $5\,764$ & 4.18 & +0.23 & $-0.08_{\pm0.19}$ & 3 & $-0.13_{\pm0.14}$ & 3 & $-0.11_{\pm0.41}$ & 4 & $-0.18_{\pm0.43}$ & 2 & H \\
HD 22879       & $5\,680$ & 3.82 & -1.03 & $+0.08_{\pm0.09}$ & 2 & $+0.34_{\pm0.18}$ & 2 & $+0.46_{\pm0.22}$ & 2 & $+0.48_{\pm0.28}$ & 2 & H \\
Sun            & $5\,707$ & 4.33 & -0.04 & $-0.07_{\pm0.07}$ & 3 & $+0.04_{\pm0.19}$ & 4 & $-0.06_{\pm0.29}$ & 5 & $+0.04_{\pm0.20}$ & 2 & F \\
$\tau$ Cet     & $5\,262$ & 4.45 & -0.56 & $-0.04_{\pm0.06}$ & 2 & $+0.47_{\pm0.36}$ & 3 & $-0.01_{\pm0.24}$ & 3 & $+0.22_{\pm0.45}$ & 1 & F \\
$\alpha$ Cen B & $5\,151$ & 4.41 & +0.18 & $-0.11_{\pm0.23}$ & 3 & $-0.03_{\pm0.03}$ & 3 & $-0.05_{\pm0.45}$ & 4 & $<+0.40$          & 1 & H \\
18 Sco         & $5\,796$ & 4.34 & +0.04 & $+0.03_{\pm0.06}$ & 3 & $+0.11_{\pm0.33}$ & 4 & $-0.11_{\pm0.32}$ & 5 & $+0.22_{\pm0.20}$ & 2 & H \\
$\mu$ Ara      & $5\,789$ & 4.39 & +0.25 & $+0.02_{\pm0.09}$ & 3 & $+0.10_{\pm0.27}$ & 3 & $-0.06_{\pm0.62}$ & 3 & $-0.10_{\pm0.12}$ & 2 & F \\
$\beta$ Vir    & $6\,061$ & 3.86 & +0.17 & $-0.07_{\pm0.19}$ & 3 & $-0.22_{\pm0.45}$ & 3 & $-0.36_{\pm0.37}$ & 4 & $-0.17_{\pm0.13}$ & 2 & H \\
\hline
\bf{Red Giants}&          &      &       &                   &   &                   &   &                   &   &                   &   &   \\
Arcturus       & $4\,286$ & 1.81 & -0.53 & $-$               & 0 & $+0.21_{\pm0.15}$ & 2 & $+0.33_{\pm0.27}$ & 3 & $+0.19_{\pm0.22}$ & 2 & U \\
$\epsilon$ Vir & $5\,197$ & 2.98 & +0.12 & $+0.77_{\pm0.19}$ & 3 & $+0.13_{\pm0.17}$ & 3 & $-0.02_{\pm0.41}$ & 3 & $+0.31_{\pm0.11}$ & 2 & F \\
$\alpha$ Tau   & $3\,839$ & 1.12 & -0.03 & $<+0.40$          & 1 & $<+0.40$          & 1 & $<+0.40$          & 1 & $<+0.40$          & 1 & F \\
\hline
\bf{K Dwarfs}  &          &      &       &                   &   &                   &   &                   &   &                   &   &   \\
$\epsilon$ Eri & $5\,170$ & 4.72 & -0.06 & $+0.23_{\pm0.17}$ & 3 & $+0.25_{\pm0.42}$ & 3 & $-$               & 0 & $+0.34_{\pm0.51}$ & 2 & F \\
\hline
\end{tabular}
\caption{\label{table_benchmarks}Ba, Eu, Gd, and Dy abundances and errors for the 19 Gaia Benchmarks stars identified 
in our sample. The number of spectral lines used to derive the \emph{r-} and \emph{s-} process abundances is indicated by $N_l$. 
The spectrograph is also indicated in the last column. We note that for the Sun we averaged parameters 
and abundances of 22 solar spectra.}
\end{table*}

\subsection{Comparison with literature Eu and Ba abundances}

We  compare our Eu and Ba abundances with recent studies;  not enough data have been published 
for Gd and Dy. The samples presented above contain 
183 HARPS stars in common with \citet{delgado_mena_2017}. These 183 stars are subgiant and dwarf stars, 
covering the domains within $4\,500<\tef<6\,200\,$K, $3.7<\logg<4.7\,$\cms~and $-0.92<\mh<+0.31\,$dex. 
The mean difference and dispersion in the adopted $\tef$, $\logg$ and $\mh$ between the two groups are about 
$\{-24;42\}\,$K, $\{-0.08;0.14\}\,$\cms~and $\{-0.04;0.04\}\,$dex, respectively. Our samples also contain 48 stars in common with 
\citet{battistini_bensby_2016}. Also subgiant and dwarfs, these stars cover the atmospheric parameter 
domains within $5\,300<\tef<6\,000\,$K, $3.8<\logg<4.6\,$\cms~and $-0.92<\mh<+0.34\,$dex. The mean difference and 
dispersion in the adopted $\tef$, $\logg,$ and $\mh$ between their study and our is about $\{-12;49\}\,$K 
in $\tef$, $\{-0.11;0.15\}\,$\cms~in~$\logg$ and $\{-0.04;0.05\}\,$dex in~$\mh$.

\begin{figure*}
\centering
\includegraphics[width=1.0\linewidth]{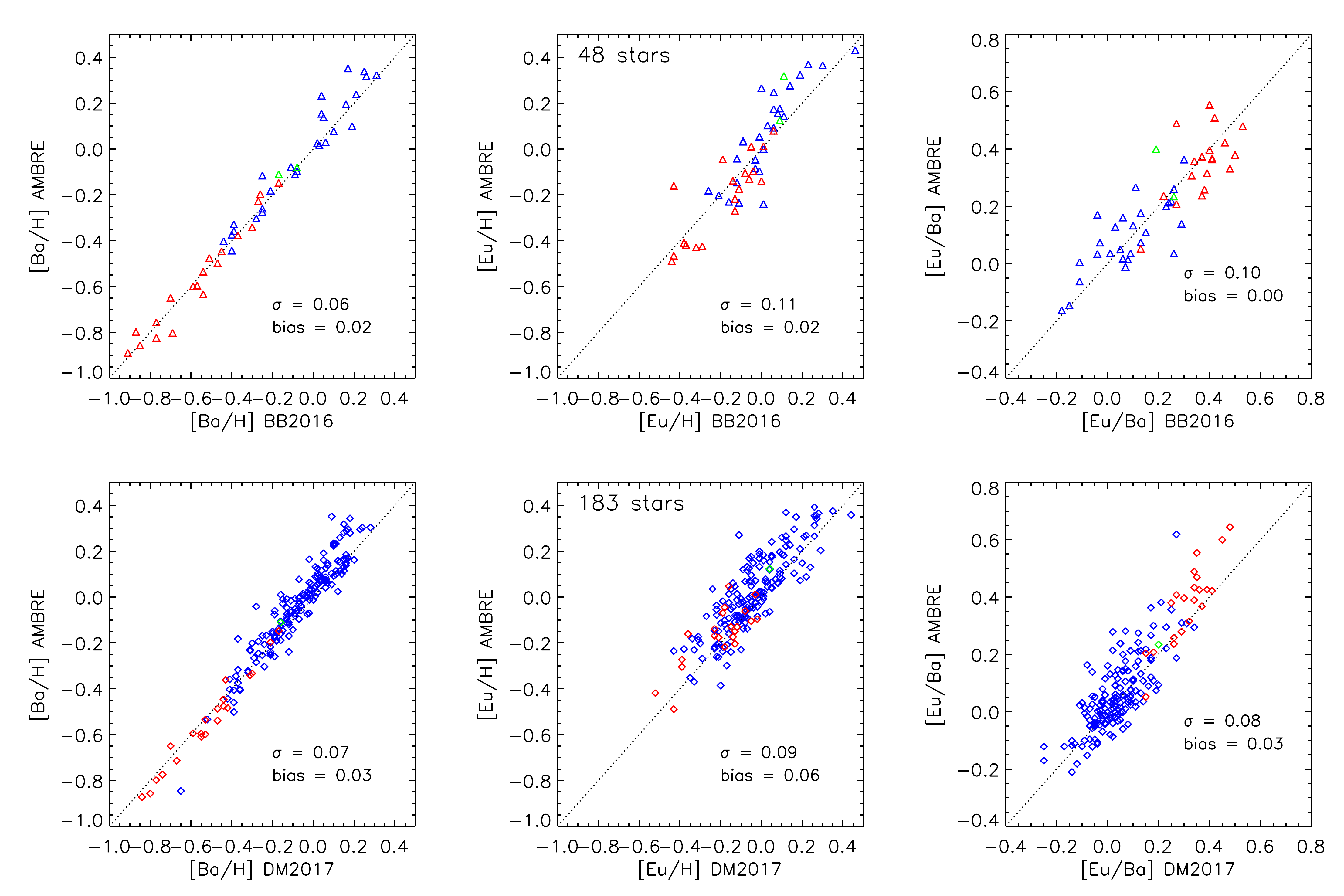}
\caption{\label{compar_delgado_bb}Comparison of AMBRE [Ba/H], [Eu/H], and [Eu/Ba] 
with those from \citet{battistini_bensby_2016} (BB2016, $\vartriangle$, top row), and \citet{delgado_mena_2017} 
(DM2017, $\diamond$, bottom row). The thin disc, thick disc, and $mr\alpha r$ stars are shown in blue,  red, and green, respectively.}
\end{figure*}

In \figurename~\ref{compar_delgado_bb}, we present comparisons between AMBRE [Ba/H], [Eu/H], and [Eu/Ba] 
and those reported by \citet{battistini_bensby_2016} and \citet{delgado_mena_2017}. We first see that the 
[Ba/H] ratio provided by AMBRE is in a very good agreement with both reference samples, showing no biases 
and weak dispersions (0.06 and 0.07 dex, respectively). We also note that the thin and the thick disc stars 
show the same trend in these comparisons (see Sect.~\ref{thin_to_thick_section} for our definition of the 
thin/thick disc labelling). Concerning [Eu/H], the comparisons with respect to \citet{battistini_bensby_2016} also shows 
 very good agreement, with no bias and low dispersion ($\sigma=0.11\,$dex). With respect to \citet{delgado_mena_2017}, 
[Eu/H] is in  good agreement as well,  also with a small dispersion and bias (bias=0.06 dex and $\sigma=0.09\,$dex). 
The biases and dispersions measured here can originate from differences in the abundance determination method, spectral line selection or 
normalization procedure. We finally compare the [Eu/Ba] ratios measured in AMBRE with both reference samples. 
We clearly see that the agreement is very good, in both the thin and the thick discs, a low dispersion being 
observed (0.10/0.08 dex with respect to the two studies). For all these chemical species, the dispersions 
between the literature values and our values are always smaller than our reported errors, confirming the good quality of 
our fully automatic analysis performed for a much larger number of stars.

\section{Selecting our working sample}\label{working_sample}

In this section, we explain how we chemically characterized the thin and thick disc stars. 
We then present our final working subsample for which we selected the best chemical abundances 
by making proper cuts in the error distributions.

We first note that in this study we only  focus 
 on dwarfs and subgiants, selecting stars with $\logg>3.5\,$\cms. In this way we do not 
expect large systematics in the abundances due to different spectral diagnostics available for 
different type of stars. Indeed, the calibration of the abundances was based on the solar 
abundances. Since barium lines, for example at 6496\AA, can be saturated for cool and metal-rich 
giants, the abundance determinations could thus suffer from larger errors. In general, cool 
stars exhibit more blended lines due to the increasing contribution of molecules, so chemical 
abundance determinations could also be  challenging and lead to high systematics, or only upper 
limits can be measurable. A solution would be to independently calibrate giants with Arcturus 
\emph{r-} and  \emph{s-}process individual abundances; however,  this is beyond  the scope of this paper and 
we note that most of our sample is dominated by dwarfs.

\subsection{Thin to thick disc dichotomy in the solar neighbourhood}\label{thin_to_thick_section}

The high quality of the statistics and homogeneity of the abundances derived in this paper thanks to the large 
HARPS, UVES, and FEROS samples allow us to study the evolution of \emph{r-}process elements in 
the two main components of the Milky Way disc: the thin disc (characterized by a low-$\alpha$ 
sequence) and the thick disc (characterized by a higher $\alpha$ sequence). It is the first 
time that Gd and Dy abundances patterns have been presented in these two Milky Way components. To this end we 
first needed to define which star belongs to each disc. We took advantage of the $\alffe$ and the 
metallicity provided by the AMBRE project. These ratios are commonly used to  chemically disentangle
the thin and the thick discs \citep{Adibekyan2011, ges_disc_recio_2014}. In this context, we followed 
the same procedure as in \citet{guiglion_2016}, applying the same chemical separation (see their Fig. 10). The main reason is that the sample presented here and the one from \citet{guiglion_2016} 
are built from similar UVES, HARPS, and FEROS samples. Metal-rich $\alpha$-rich stars (mr$\alpha$r) with 
$\mh>-0.15\,$dex and $\alffe$ above the separation are also treated separately because they are too 
metal-rich compared to the classical definition of the thick disc.

\begin{figure}
\centering
\includegraphics[width=1.0\linewidth]{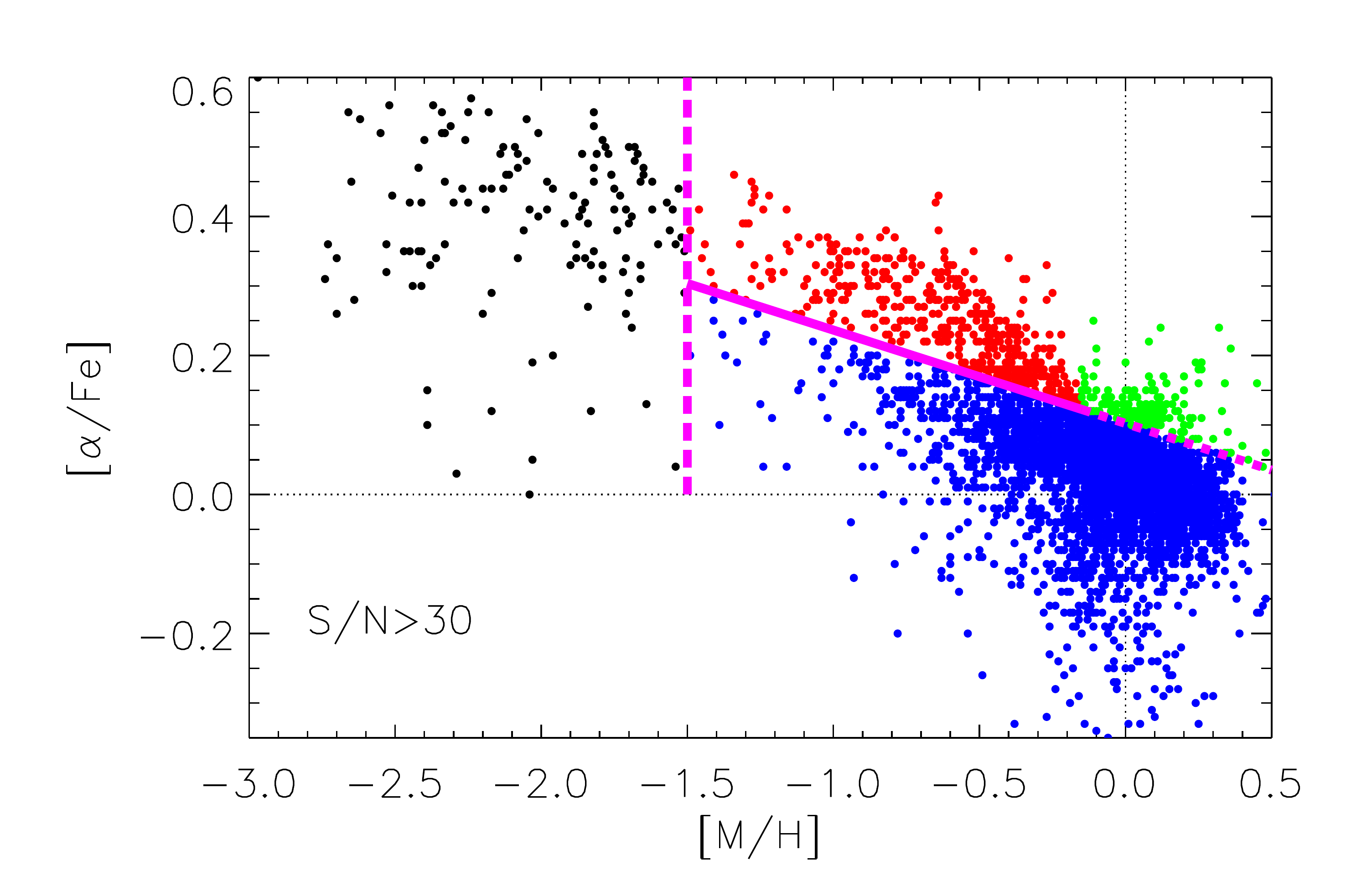}
\caption{\label{thin_to_thick} $\alffe$  as a function of $\mh$ for a subsample of stars 
with $\snr>30$. The full magenta line shows the thin to thick disc separation. The magenta dashed 
line shows the extrapolated separation for $\mh>-0.15\,$dex, while the vertical  long-dashed line 
characterizes halo stars ($\mh<-1.25\,$dex). The thin disc stars are colour-coded in blue, while thick 
disc members are in red. The metal-rich $\alpha$-rich stars are shown in green, and halo stars in black.}
\end{figure}

We first selected a high $\snr$ subsample from the $5\,057$ stars of the Ba sample, as seen in 
\figurename~\ref{thin_to_thick}. Chemically characterized with $\alffe$ and $\mh$, we present the resulting three populations 
of disc stars: the thin disc, the thick disc, and metal-rich $\alpha$-rich stars. 
The abundance pattern presented in \figurename~\ref{thin_to_thick} is consistent when using the 
three other samples of \tablename~\ref{nb_star_elem_spectro} ($6\,268$, $5\,431$, $5\,479$ stars, respectively) 
and we adopted the magenta line of \figurename~\ref{thin_to_thick} 
to tag these stars still using the $\alffe$ versus $\mh$ plane.\\
We flagged stars with $\mh<-1.50\,$dex 
as halo stars. We note that these halo stars present weak spectral lines and larger errors, so we 
do not expect to include many of them in our final working sample (see Sect~\ref{best_abund}). 
Additionally, $95\%$ of halo targets were observed with UVES or FEROS, at $R=40\,000$, making the 
detection of such weak lines more difficult. We finally note that our sample contains (labelled) thin disc stars with $\mh<-0.7\,$dex 
characterized by a lower-$\alpha$ content with respect to thick disc stars in the same metallicity range. 
We are conscious that in the metal-poor regime a small contamination by halo stars might exist.

\subsection{Selecting the best abundances}\label{best_abund}

In order to understand the evolution of Ba, Eu, Gd, and Dy in the Milky Way discs, we selected the 
best chemical abundances among the samples presented in \tablename~\ref{nb_star_elem_spectro}. We 
note that we were not able to derive Ba, Eu, Gd, and Dy for all of these stars, or in some cases were  able to  provide 
 only upper limits, which we do not consider here. For a given element, we then removed 
 16\% (for Ba), 26\% (for Eu), 35\% (for Gd), and 43\% (for Dy) of the targets. We also 
selected abundances derived with at least two detectable spectral lines for a given element.

\begin{figure}
\centering
\includegraphics[width=1.0\linewidth]{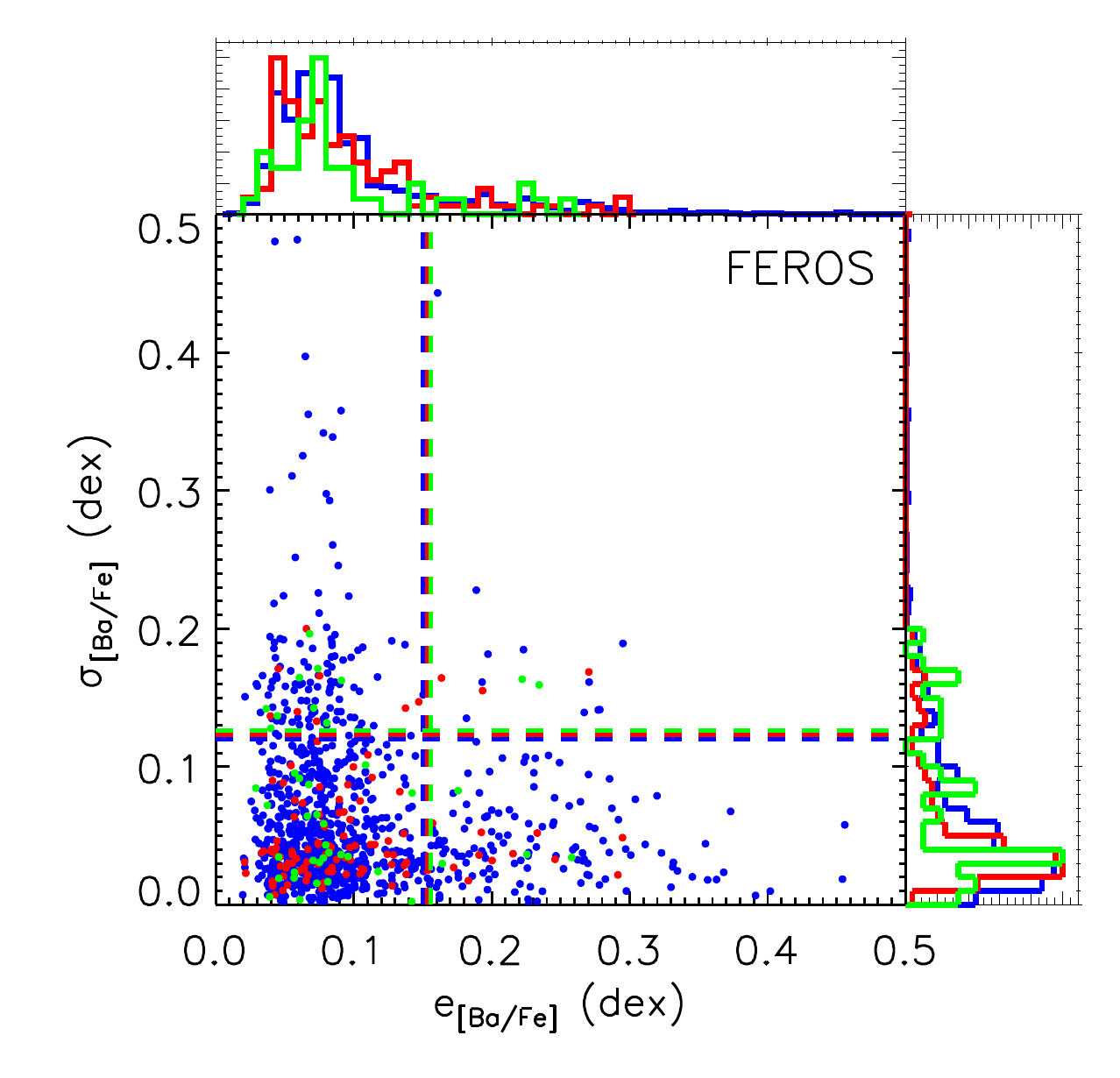}
\caption{\label{feros_eba_thin_thick_mrar} For the Ba abundance, line-to-line scatter as a function of the error 
due to the atmospheric parameters for the FEROS sample. The thin disc (blue dots, 917 stars),  thick disc 
(red dots, 112 stars), and  metal-rich $\alpha$-rich (green dots, 43 stars) samples are shown, with their corresponding 
normalized distributions. The dashed lines show the adopted  cut in errors.}
\end{figure}

To better clean our samples, we then took advantage of the error budget. For a given spectrograph 
and a given element, we carefully visualized the distributions of  the error due to the atmospheric 
parameters, and of the line-to-line scatter. In \figurename~\ref{feros_eba_thin_thick_mrar} we show how 
we proceeded. For the Ba measurements in the FEROS sample, we cut the tails of the distributions, 
in this case 0.12 dex for $\sigma_{[\text{Ba/Fe}]}$ and 0.15 for $\text{e}_{[\text{Ba/Fe}]}$, for the 
three subpopulations. These cuts are typical for this example, but depends on the elements and the 
population considered. We repeated this operation for the four elements Ba, Eu, Gd, and Dy, and for the 
three samples (FEROS, UVES, HARPS). We also recall  that the HARPS analysis was performed at $R=110\,000$, 
while that of FEROS and UVES  was done at $R=40\,000$. We then applied a different cut in errors for a given 
sample, HARPS providing generally more accurate abundances.

The final samples are presented in \tablename~\ref{nb_star_final_sample} and \tablename~\ref{selection_erros}, 
where the number of stars in each sample is shown and the mean abundance error is provided. We unfortunately 
have no star belonging to the mr$\alpha$r population in the UVES sample, nor halo stars with reliable abundances. 
The typical total errors are 0.1 dex in Ba, 0.15 dex in Eu, and 0.2 dex in Gd and Dy.

\begin{table}
\centering
\begin{tabular}[c]{c|c c c c}
$N_{stars}$ & Thin & Thick & mr$\alpha$r & sample \\
\hline
\hline
Ba &          740 &           96 &           37 & FEROS \\
 &          726 &           74 &            5 & HARPS \\
 &           10 &            6 & - & UVES \\
 & \bf{        1476} & \bf{         176} & \bf{          42} & $\sum$ \\
\hline
Eu &          501 &           69 &            9 & FEROS \\
 &          586 &           66 &            2 & HARPS \\
 &          144 &           36 & - & UVES \\
 & \bf{        1231} & \bf{         171} & \bf{          11} & $\sum$ \\
\hline
Gd &          403 &           52 &           11 & FEROS \\
 &          443 &           39 &            4 & HARPS\\
 &           27 &            2 & - & UVES \\
 & \bf{         873} & \bf{          93} & \bf{          15} & $\sum$ \\
\hline
Dy &          342 &           38 &           12 & FEROS \\
 &          560 &           54 &            5 & HARPS \\
 &           42 &            2 & - & UVES \\
 & \bf{         944} & \bf{          94} & \bf{          17} & $\sum$ \\
\hline

\end{tabular}
\caption{\label{nb_star_final_sample}Number of selected dwarf stars with available best Ba, Eu, Gd, and Dy 
abundances for each sample (FEROS, HARPS, UVES) and each population (thin/thick discs, and mr$\alpha$r).}
\end{table}

\begin{table}
\centering
\begin{tabular}[c]{c|c|c c c|c}
Elem. & error & Thin & Thick & $\alpha$rmr & sample \\
\hline
\hline
[Ba/Fe] & $\langle e \rangle$ & 0.07 & 0.07 & 0.07 & FEROS \\
 & & 0.07 & 0.07 & 0.08 & HARPS \\
 & & 0.06 & 0.04 &  -   & UVES \\
 & $\langle \sigma \rangle$ & 0.06 & 0.05 & 0.07 & FEROS \\
 & & 0.05 & 0.04 & 0.04 & HARPS \\
 & & 0.11 & 0.06 &  -   & UVES \\
 & $\langle e_{tot} \rangle$ & 0.10 & 0.09 & 0.11 & FEROS \\
 & & 0.09 & 0.08 & 0.10 & HARPS \\
 & & 0.13 & 0.07 &  -   & UVES \\
\hline
[Eu/Fe] & $\langle e \rangle$ & 0.11 & 0.11 & 0.11 & FEROS \\
 & & 0.12 & 0.12 & 0.09 & HARPS \\
 & & 0.10 & 0.10 &  -   & UVES \\
 & $\langle \sigma \rangle$ & 0.12 & 0.12 & 0.14 & FEROS \\
 & & 0.13 & 0.13 & 0.21 & HARPS \\
 & & 0.06 & 0.12 &  -   & UVES \\
 & $\langle e_{tot} \rangle$ & 0.18 & 0.17 & 0.18 & FEROS \\
 & & 0.18 & 0.18 & 0.23 & HARPS \\
 & & 0.13 & 0.18 &  -   & UVES \\
\hline
[Gd/Fe] & $\langle e \rangle$ & 0.16 & 0.14 & 0.16 & FEROS \\
 & & 0.14 & 0.14 & 0.12 & HARPS \\
 & & 0.13 & 0.08 &  -   & UVES \\
 & $\langle \sigma \rangle$ & 0.14 & 0.18 & 0.11 & FEROS \\
 & & 0.12 & 0.14 & 0.10 & HARPS \\
 & & 0.25 & 0.24 &  -   & UVES \\
 & $\langle e_{tot} \rangle$ & 0.22 & 0.24 & 0.21 & FEROS \\
 & & 0.19 & 0.21 & 0.16 & HARPS \\
 & & 0.29 & 0.26 &  -   & UVES \\
\hline
[Dy/Fe] & $\langle e \rangle$ & 0.15 & 0.15 & 0.14 & FEROS \\
 & & 0.14 & 0.13 & 0.18 & HARPS \\
 & & 0.14 & 0.09 &  -   & UVES \\
 & $\langle \sigma \rangle$ & 0.08 & 0.10 & 0.07 & FEROS \\
 & & 0.07 & 0.09 & 0.08 & HARPS \\
 & & 0.11 & 0.06 &  -   & UVES \\
 & $\langle e_{tot} \rangle$ & 0.18 & 0.19 & 0.17 & FEROS \\
 & & 0.16 & 0.17 & 0.20 & HARPS \\
 & & 0.19 & 0.12 &  -   & UVES \\
\hline
\end{tabular}
\caption{\label{selection_erros}Mean error ($e$, due to atmospheric parameter errors), 
mean line-to-line scatter ($\sigma$), and mean total error (defined as $e_{tot}=\sqrt{e^2+\sigma^2}$) 
for each dwarf subsample (FEROS, HARPS, UVES) and each population (thin/thick discs, and mr$\alpha$r).}
\end{table}

\begin{figure}
\centering
\includegraphics[width=1.0\linewidth]{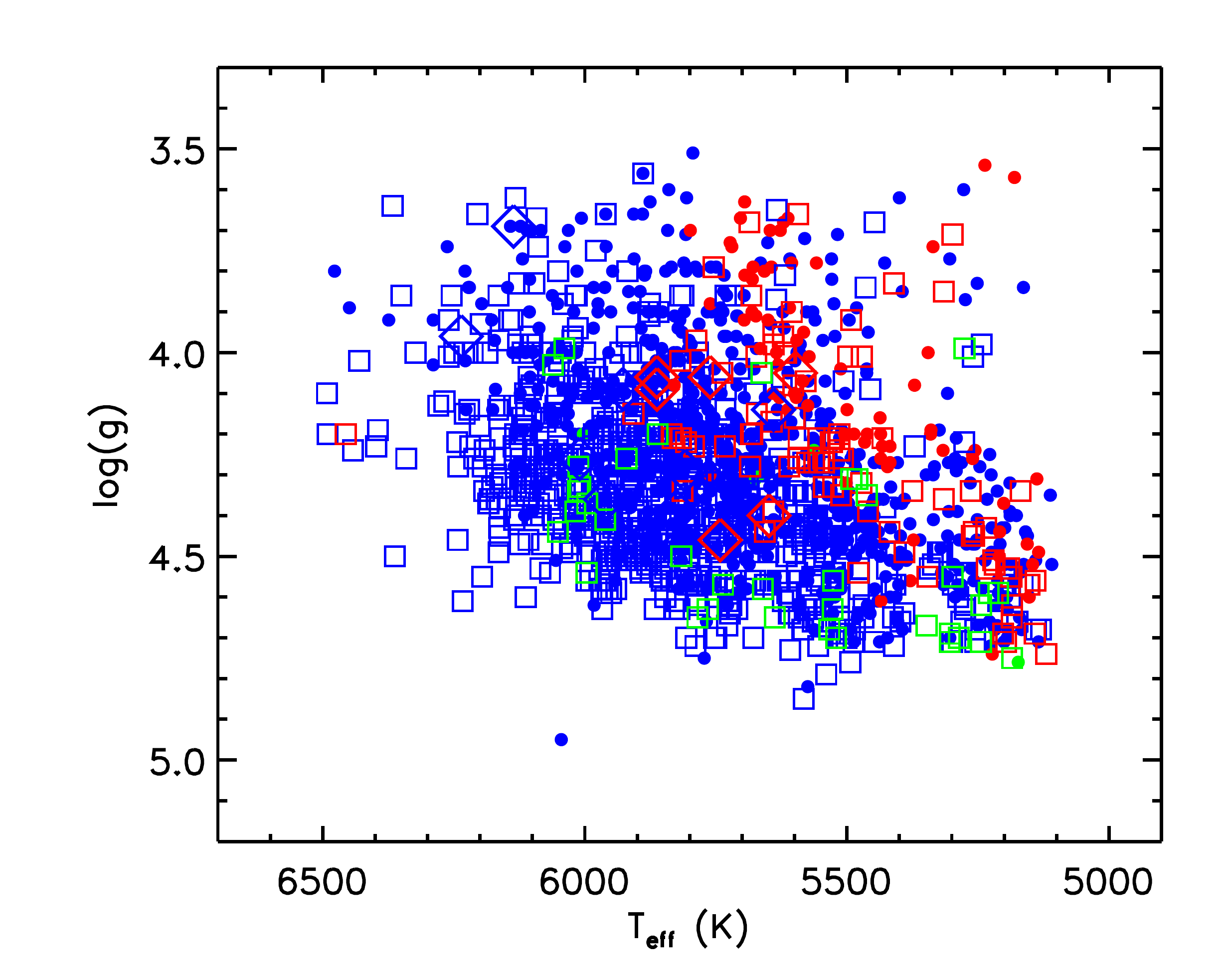}
\caption{\label{teff_logg_final_ba}$\tef\,vs.\,\logg$ for the Ba sample. The thin disc (blue, 1476 stars), 
 thick disc (red, 176 stars), and  metal-rich $\alpha$-rich (green, 42 stars) samples are shown. Different 
symbols are used for HARPS (\textbullet), FEROS ($\square$), and UVES ($\diamond$). We note that the coverage 
on this $\tef\,vs.\,\logg$ plane is similar to the Eu, Gd, and Dy sample, in addition to a different number of stars.}
\end{figure}

We note that our final samples cover the atmospheric parameter domains $5\,100<\tef<6\,300\,$K 
and $3.5<\logg<5.0$, which is  illustrated in \figurename~\ref{teff_logg_final_ba} for the Ba sample. 
We do not have stars cooler than $5\,100\,$K because of large errors, and these stars have been 
removed when applying cuts of the errors. Finally, we point out that \citet{korotin_2011} reported 
that Ba abundances can suffer from NLTE effects for hot stars. Our thin disc stars with $\tef<6\,100\,$K 
show on average slightly lower Ba abundances ($-0.06\,$dex) over the whole metallicity range in both 
thin and thick discs; Ba abundances are fully consistent between hot and cool stars. No such hot stars 
are present in the mr$\alpha$r population. We then decide to keep these hot stars with their LTE 
abundances in the present samples.

\section{The \emph{r-}process element evolution in the Milky Way}\label{discussion}

\subsection{Ba and \emph{r-}process abundances trends with metallicity}

\begin{figure*}
\centering
\includegraphics[width=1.0\linewidth]{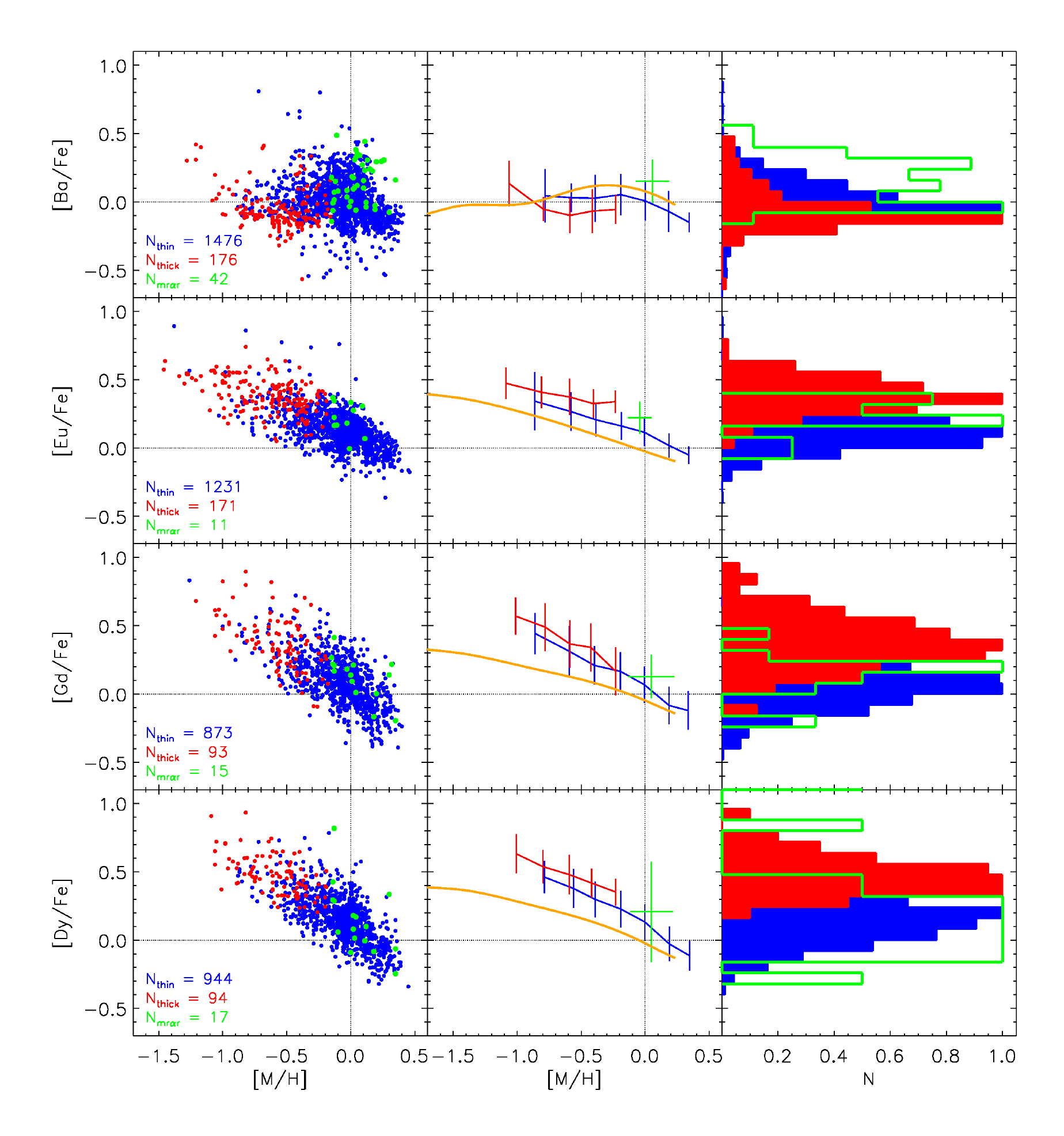}
\caption{\label{working_sample_ba_eu_gd_dy}\emph{Left:} [Ba/Fe], [Eu/Fe], [Gd/Fe],
and [Dy/Fe] ratios in the thin disc (blue), thick disc (red), and in the $\text{mr}\alpha\text{r}$ 
population (green). \emph{Middle}: Average abundances binned in metallicity every 0.2 dex. The 
orange curves show the Galactic chemical evolution models of \citet{prantzos_2018}. \emph{Right}: 
Corresponding normalized distributions.}
\end{figure*}

We present chemical abundance trends of Ba and \emph{r-}process in 
\figurename~\ref{working_sample_ba_eu_gd_dy}, for the thin disc, thick disc, and the 
$\text{mr}\alpha\text{r}$ populations, using the working sample we defined above. We 
computed average trends (and their associated standard deviations) of Ba, Eu, Gd, and 
Dy as a function of the metallicity, in the thin and thick discs and for the 
$\text{mr}\alpha\text{r}$ population (middle panel). We used a typical metallicity bin 
of 0.2 dex, and a single bin for stars with $\mh<-1.0$, due to lower statistics. In the 
same way, we adopted a single bin for the $\text{mr}\alpha\text{r}$ population. We checked 
that the trends presented here are robust when changing the metallicity binning, typically 
by a shift of 0.05 dex. The typical number of stars per bin is 100--200 for the thin disc 
and 20--30 for the thick disc. We also show histogram distributions. We note that our 
separation between thin disc, thick disc, and $\text{mr}\alpha\text{r}$ stars is purely 
based on the chemistry ($\alffe$ versus $\,\mh$), and we cannot exclude any possible contamination 
between these three populations, especially at high metallicity.

- The Ba abundance in the thin disc tends to be constant from the metal-poor regime 
($\mh\sim-1\,$dex) to solar metallicity, and then decreases for super-solar $\mh$ 
revealing a higher rate production of Fe than Ba in the disc at recent epochs. The 
scatter seems to be the largest around solar $\mh$, while this dispersion reduces when 
going towards metal-poor and metal-rich regimes. The bulk of the thin disc shows roughly 
solar [Ba/Fe] ratios. This trend is consistent with previous [Ba/Fe] patterns from the 
literature \citep{battistini_bensby_2016, delgado_mena_2017}. Our thin disc data also seem  
to be consistent with the prediction of \citet{bisterzo_2017}, in addition to a delay in the maximum 
[Ba/Fe] ratio. We recall that their Galactic chemical evolution model is based on a three-zone 
model (thin and thick discs + halo), with two main processes: a primary \emph{r-}process 
production in the Galaxy from moderately massive Type II supernovae ($8-10\,M_{\bigodot}$, 
and a second \emph{s-}process by low- and intermediate-mass  AGB stars.

We also note  the presence of thin disc stars with peculiar Ba abundances ([Ba/Fe]>0.5 or 
[Ba/Fe]<-0.5 dex), especially for $\mh<-0.2\,$dex, that could be interpreted as contamination by halo 
stars \citep{suda_2011}.

The thick disc is characterized by a flat sequence around [Ba/Fe]$\sim-0.15\,$dex, with a quite 
constant dispersion with $\mh$ and then an increase at $\mh<-0.8\,$dex, probably caused by a 
contamination by halo stars. We note that the thick disc clearly presents a smaller Ba abundance 
with respect to the thin disc, in the same metallicity range, and that both discs 
show the same dispersion $(\sigma_{[\text{Ba}/\text{Fe}]}=0.15\,$dex). In addition, our thick 
disc data do not match the [Ba/Fe] model of \citet{bisterzo_2017}, predicting an increase in 
[Ba/Fe] as a function of $\mh$. \citet{delgado_mena_2017} also observed a flat trend in their 
data, and evoked the fact that a too weak metallicity coverage of their data could create such 
a mismatch, especially for $\mh<-0.8\,$dex. In our data, we cover a wider metallicity range, but 
the \citet{bisterzo_2017} predictions still do not fit our observations in the thick disc.

The $\text{mr}\alpha\text{r}$ population Ba abundance seems to be consistent with the thin disc 
pattern, even though it is  a bit more Ba-rich. 

- The [Eu/Fe] ratio in the thin disc decreases in lower metallicity stars, typically from 
+0.4/0.5 dex at $\mh\sim-1.0\,$dex to $[\text{Eu}/\text{Fe}]=+0.1\,$dex at $\mh=0$, and 
solar value for $\mh>0$, with a distribution peaking at [Eu/Fe]$\sim+0.1\,$dex. The thick 
disc Eu abundance also follows  a decreasing sequence with increasing metallicity, showing a 
continuous sequence with the thin disc, peaking at [Eu/Fe]$\sim+0.35\,$dex. On the same 
metallicity domain ($\mh<-0.15\,$dex) the thick disc is  more [Eu/Fe]-rich by about 
[Eu/Fe]$=+0.17\,$dex. These two observations are consistent with Galactic chemical evolution 
model predictions from \citet{bisterzo_2017}. Both thin and thick discs show the same scatter 
in their [Eu/Fe] pattern ($\sigma_{[\text{Eu}/\text{Fe}]}=0.13\,$dex). We also note that 
[Eu/Fe] shows a typical $\alffe$  behaviour in both discs, consistent with a  lower 
production of Eu with time while the Fe production increases. We also clearly show that the 
thick disc is enriched in \emph{r-}process with respect to the thin disc, when using binned 
data. The $\text{mr}\alpha\text{r}$ population [Eu/Fe] ratio seems here to be consistent with 
the thin disc pattern. We also note here  the presence of peculiar stars with [Eu/Fe]>+0.7 dex 
in the thin disc, also showing high [Gd/Fe] and [Dy/Fe]. These stars, with low-$\alffe$ pattern, 
were also characterized as thin disc members by \citet{mikolaitis_2017}.

- [Gd/Fe] and [Dy/Fe] show very similar patterns in the three Milky Way disc populations and are 
almost consistent with the behaviour of [Eu/Fe]. In the thin disc, the [Gd/Fe] and [Dy/Fe] 
ratios decrease  from +0.5/0.6 dex at $\mh\sim-0.8\,$dex, to 
$[\text{Eu}/\text{Fe}]\sim+0.1\,$dex at $\mh=0$, and reach -0.2 dex at super-solar $\mh$, 
with their distribution peaking at $[\text{Gd}/\text{Fe}]\sim+0.15\,$dex. The [Gd/Fe] and 
[Dy/Fe] histograms both show the same dispersion ($\sigma=0.17\,$dex). The thick disc, as for 
[Eu/Fe], shows a continuous sequence with the thin disc, reaching  [Gd/Fe] and [Dy/Fe] 
$\sim0.7/0.7\,$dex at $\mh\sim-1\,$dex. The thick disc [Gd/Fe] and [Dy/Fe] ratios both peak at 
$\sim+0.40\,$dex, but [Gd/Fe] shows higher dispersion with respect to [Dy/Fe] 
($\sigma_{[\text{Gd}/\text{Fe}]}=0.20\,$dex against $\sigma_{[\text{Dy}/\text{Fe}]}=0.14\,$dex). 
The [Gd/Fe] and [Dy/Fe] ratios are characterized by a steeper decrease as a function of the 
metallicity in both discs than [Eu/Fe], with a higher dispersion, especially for [Gd/Fe]. Here the 
[Gd/Fe] and [Dy/Fe] patterns of the $\text{mr}\alpha\text{r}$ population also seem   to be 
consistent  with thin disc patterns. The $\text{mr}\alpha\text{r}$ population also tends  to be 
slightly more enriched in \emph{r-}process than  the thin disc, and seems to be in the continuity 
of the thick disc. These stars are then both $\alpha$-rich and \emph{r-}rich.

There are no models of Galactic chemical evolution directly comparable
to our data, i.e. considering explicitly the cases of the thin and
thick discs and including the elements heavier than Fe. This has been
done only for elements up to the Fe-peak (e.g. \citealt{minchev_2013},
\citealt{kubryk_2015}, see a comparison of the latter with AMBRE data in
\citealt{mikolaitis_2017}). To date the evolution of heavier elements has been
studied  only with one-zone models (i.e. \citealt{travaglio_2004}).
The most complete model is the one recently published by \citet{prantzos_2018}. 
It includes all elements up to U and their isotopes; a complete set of 
metallicity-dependent yields of massive, rotating
stars from \citet{limongi_chieffi} (including the weak \emph{s-}process)
and of low- and intermediate-mass stars
(including the main \emph{s-}process); as well as a fiducial \emph{r-}component from
massive stars for all isotopes, calibrated to the corresponding yield of
${}^{16}\text{O}$. In this way, all heavy elements and isotopes are found to be well
co-produced at their corresponding solar values and at the time of the
solar system formation, 4.5 Gy ago (see figures 10 and 12 in that
paper). The local evolution of several elements, i.e. the behaviour of
[x/Fe] versus  [Fe/H ]) is also found to be well reproduced when compared to
observations (their fig. 16); however, the adopted data sets are not
homogenized and no distinction is made between thin and thick discs
(neither in the model, nor in the data), making it difficult to draw
significant conclusions.

In \figurename~\ref{working_sample_ba_eu_gd_dy}, we compare our data to the 
aforementioned results of \citet{prantzos_2018} (orange curve). 
The model curves lie below our data, even for the thin
disc. This is obviously because our data display a
super-solar [r/Fe]=0.1--0.15 dex at [Fe/H]=0, a fact impossible to
reproduce by any one-zone model: such models are meant to produce a solar
pattern for all elements 4.5 Gy ago. The interpretation of our data
requires dedicated multi-zone models, either semi-analytical or fully
chemo-dynamical. In particular, the role of neutron-star mergers (NSM)
in the production of \emph{r-}elements should be considered in such models
after the recent joint detection of electromagnetic and gravitational
signals from the gamma-ray burst GW170817/GRB170817A (\citealt{abbott_kilo_2017},
\citealt{pian_2017} and references therein).

\subsection{Ratios of pure \emph{r-}element to barium}

\begin{figure*}
\centering
\includegraphics[width=1.0\linewidth]{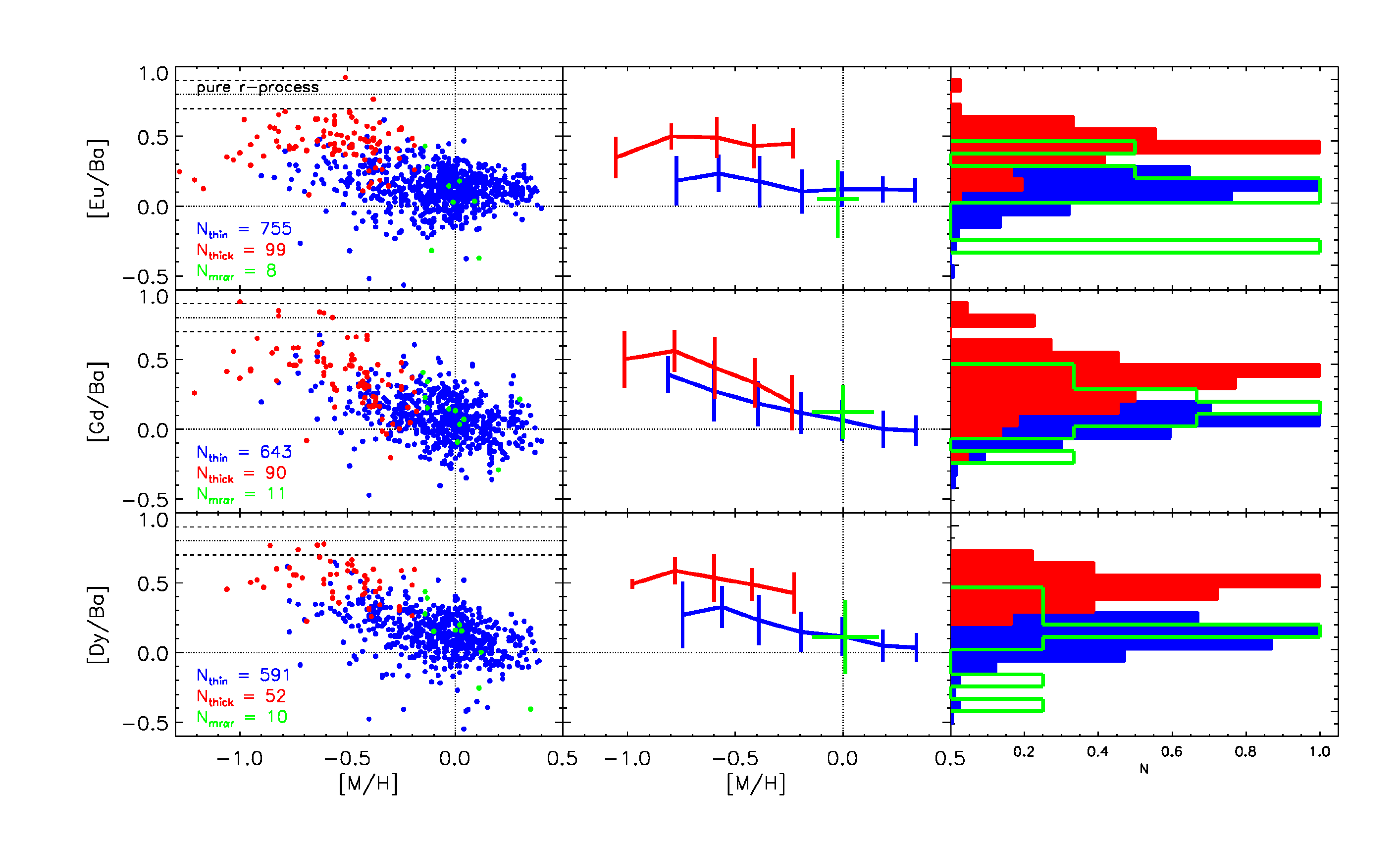}
\caption{\label{working_sample_ba_eu_gd_dy_equ_rapports}[Eu/Ba], [Gd/Ba], and [Dy/Ba] ratios in the thin disc, thick disc, and 
in the $\text{mr}\alpha\text{r}$ population (colour-coding as in \figurename~\ref{working_sample_ba_eu_gd_dy}). 
We also present average abundances binned in metallicity every 0.2 dex. The thick dotted line 
shows the pure \emph{r-}process ratio from the model of \citet{bisterzo_2014} with its 0.1 dex error. The right panel 
shows normalized distributions.}
\end{figure*}

In order to quantify the relative importance of the \emph{r-} and \emph{s-} channels 
during the evolution of the Milky Way,  we present in \figurename~\ref{working_sample_ba_eu_gd_dy_equ_rapports} 
Eu, Gd, and Dy abundances (pure \emph{r-}process elements) with respect to Ba (pure \emph{s-}process) 
as a function of the metallicity $\mh$ for the different disc components. We point out that our 
statistics becomes lower since we kept only stars with measurement of Ba and one of the \emph{r-}element.

In both thin and thick discs, our [Eu/Ba] ratio looks quite constant with $\mh$ within the error 
bars. The thick disc shows a higher [Eu/Ba] ratio ([Eu/Ba]$\sim0.45\,$dex) than the thin 
disc ([Eu/Ba]$\sim0.15\,$dex). The thick disc pattern is consistent with findings of previous studies, 
 for example \citet{battistini_bensby_2016} and \citet{delgado_mena_2017}. However, in 
\citet{delgado_mena_2017} the thin disc [Eu/Ba] pattern tends to decrease until solar value for $\mh<0$, 
and increase for $\mh>0\,$dex, in contradiction with our observations. We note that our statistics is 
higher by at least a factor of two. On the contrary, we show that the [Gd/Ba] ratio is characterized by 
a decrease in both discs, about -0.4 dex over 1 dex in $\mh$ for the thin discs, and -0.3 dex over 0.7 
dex in $\mh$ for the thick disc. Similar behaviour is also observed for [Dy/Ba] in addition to a shallower decrease revealing a 
possible different production history for Eu and Gd-Dy, as confirmed in 
\figurename~\ref{working_sample_ba_eu_gd_dy_equ_rapports_r}. We note that [Gd/Ba] and [Dy/Ba] clearly 
peak at high ratios in the thick disc ([Gd, Dy/Ba]$\sim0.45\,$dex) than in the thin disc 
([Gd, Dy/Ba]$\sim0.15\,$dex).

The $mr\alpha r$ population 
shows patterns consistent with thin disc chemistry. According to \figurename~\ref{working_sample_ba_eu_gd_dy}, 
these stars are $\alpha$-rich and \emph{r-}rich (like thick disc stars), but their Ba is very different from thick disc 
stars. This result raises one more open question on the nucleosynthesis processes history of these two families 
of elements.

Finally, the [Eu/Ba], [Gd/Ba], and [Dy/Ba] ratios are close to pure \emph{r-}process in the metal-poor regime, 
and this is a sign that at the early epoch of our Galaxy, the \emph{r-}process was the dominant neutron-capture 
process \citep{bisterzo_2014}. Then, the [r/Ba] ratios decrease when AGB stars start contributing predominantly 
to the ISM enrichment in \emph{s-}process elements. As a result, [r/Ba] ratios decrease towards the solar value 
at solar metallicity. We note that it is the first time that such trends have been presented for Gd and Dy.

\subsection{Average \emph{r-}process abundance trends}

\begin{figure}
\centering
\includegraphics[width=1.0\linewidth]{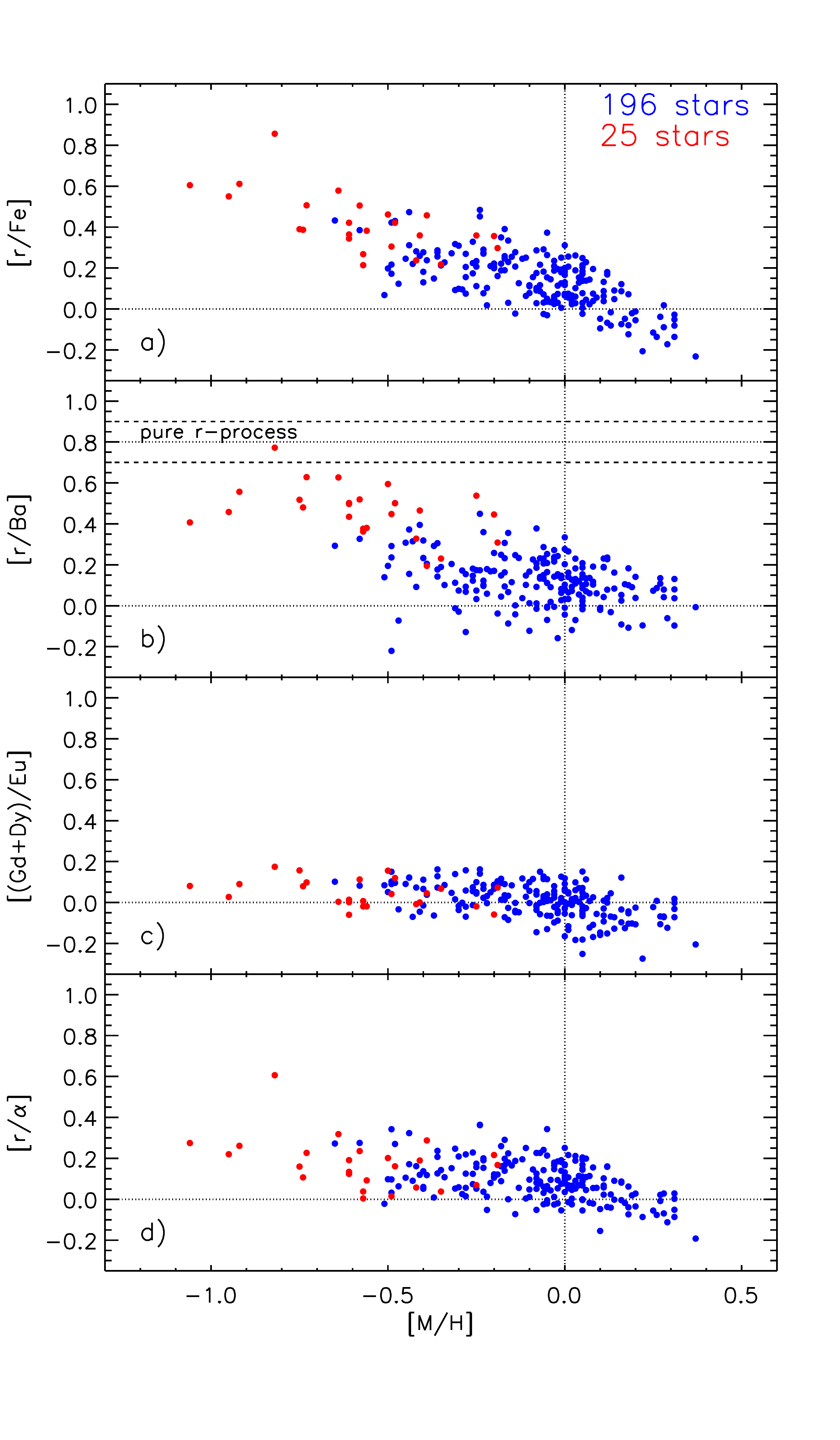}
\caption{\label{working_sample_ba_eu_gd_dy_equ_rapports_r}As a function of the metallicity: a) Average \emph{r-}process 
abundance (defined as the mean of Eu, Gd, and Dy) over Fe; b) [\emph{r}/Ba] ; c) [(Gd+Dy)/Eu] ; 
d) [\emph{r}/Ba] .  Colour-coding as in \figurename~\ref{working_sample_ba_eu_gd_dy}.}

\end{figure}

In \figurename~\ref{working_sample_ba_eu_gd_dy_equ_rapports_r}, we finally show the average 
\emph{r-}process abundance pattern (defined as the average Eu, Gd, and Dy abundances) in both 
discs as a function of the metallicity, and confirm some trends already seen in 
\figurename~\ref{working_sample_ba_eu_gd_dy_equ_rapports}. It corresponds to stars from our working 
sample with available Ba, Eu, Gd, and Dy, explaining the lower statistics in these plots. First, 
the [\emph{r}/Fe] ratio shows a narrow sequence with a small dispersion. It is clear that the thin 
and the thick discs form a continuous sequence. Moreover, on the one hand the [\emph{r}/Ba] ratio is 
globally characterized by a weakly scattered sequence decreasing from the pure \emph{r-}process 
abundance of +0.8 dex at $\mh\sim-0.8\,$dex to $[r/\text{Ba}]\sim+0.1\,$dex at $\mh=0$ (we already 
noted that $[r/\text{Fe}]>0$ in this metallicity regime). On the other hand, the thick disc is 
roughly constant at $[r/\text{Ba}]\sim+0.50\,$dex for $\mh<-0.2\,$dex, but then seems to decrease, 
while the thin disc is rather flat beyond $\mh>-0.5\,$dex.We also show that the [(Gd+Dy)/Eu] ratio is 
not constant as a function of the metallicity, revealing a possible different production history 
between Eu and Gd-Dy in both discs.

Finally, we took advantage of the $\alffe$ ratio to trace the ratio between \emph{r-}process and 
$\alpha$-elements\footnote{Using the $\mgfe$ ratio from \citet{mikolaitis_2017} 
provides the same results.}. The $[r/\alpha]$ ratio is clearly not constant as a function of the 
metallicity and tends to decrease by about +0.15 dex from $\mh\sim-1\,$dex to $[r/\alpha]\sim+0.1\,$dex 
at $\mh=0$. Here again, no clear thin to thick disc separation is visible with $[r/\alpha]$. 
Interestingly, $\alpha$-elements are predominately synthesized in Type II supernovae, while it is 
known that such core-collapse supernovae are also suitable sites for \emph{r-}process. The clear 
slope with $\mh$ indicates that supernovae of different properties contribute to the synthesis of 
\emph{r-}process elements and $\alpha$-elements, but with different efficiencies/yields. However, 
since the recent observational evidence of \emph{r-}process synthesis via neutron-stars mergers, 
this trend cannot only be explained by the role of Type II supernovae.

\section{Conclusion}\label{conclusiooooonnnn}

In this study, our goal was understanding the evolution of Milky Way disc \emph{r-}process 
abundances. We built a homogeneous catalogue of chemical abundances of Ba (pure \emph{s-}element) 
and Eu, Gd, and Dy (pure \emph{r-}elements). In the literature, such a catalogue with high 
statistics is still lacking. As a result, the chemical evolution of pure \emph{r-}process elements 
is still a matter of debate.

We took advantage of the HARPS, FEROS, and UVES ESO archives, coupled with the atmospheric 
parameters of the AMBRE project \citep{laverny_2012}. We performed an automatic derivation 
of individual chemical abundances and errors of Ba ($5\,057$ stars), Eu ($6\,268$ stars), 
Gd ($5\,431$ stars), and Dy ($5\,479$ stars) thanks to the pipeline GAUGUIN \citep{guiglion_2016}. 
It is the first time that such a homogeneous data set has been provided, especially for Gd and Dy, and 
that covers such a wide metallicity range ($-1.5<\mh<+0.5\,$dex). Comparisons of our abundances with 
previous studies show a very good agreement. We also provided such chemical abundances for 19 Gaia 
benchmark stars.

From this catalogue, we selected dwarf stars with the most accurate abundances of Ba 
($1\,694$ stars), Eu ($1\,413$ stars), Gd ($981$ stars), and Dy ($1\,055$ stars) and 
investigated the chemical abundance patterns of these four elements in the Milky Way disc,
more precisely focusing on the thin disc, the thick disc, and the metal-rich $\alpha$-rich 
population. Identifying such disc stellar populations was done using a chemical 
separation in the $\alffe$ versus $\mh$ plan. 
We summarize here our main results:\\
- The [Eu/Fe] ratio follows a continuous sequence from the thin disc to the thick disc, with respect to the metallicity.\\
- In thick disc stars, the [Eu/Ba] ratio is rather constant, while the [Gd/Ba] and [Dy/Ba] ratios decrease as a function 
of the metallicity. These observations clearly indicate a different nucleosynthesis history in the thick disc between Eu and Gd-Dy.\\
- We find that the $mr\alpha r$ population abundance patterns are consistent with the thin disc chemistry. These 
stars tend to be both enriched in $\alpha$- and \emph{r-}process elements, (like thick disc stars), but their [Ba/Fe] is very 
different from thick disc stars.\\
- We find that the [\emph{r}/Fe] ratio in the thin disc is roughly around +0.1 dex at solar metallicity, which is not the case for Ba.\\
- We also provided average [Ba, Eu, Gd, Dy/Fe] and [Eu, Gd, Dy/Ba] ratios as a function of the metallicity, 
with associated dispersion. This data is crucial when one wants to constrain Galactic chemical evolution model, more 
particularly on the stellar yields.\\
- We compared our data with the last model of \citet{prantzos_2018} that includes 
yields of rotating massive stars. In addition to the fact that it is a one-zone model, we find a good quantitative match for [Ba/Fe]. 
For [Eu, Gd, Dy/Fe] the model underpredicts the observations, being calibrated to obtain [\emph{r}/Fe]=0 at [Fe/H]=0. 
Taken at face value, the observations imply that the average stellar [\emph{r}/Fe] of the disc is super-solar at [Fe/H]=0, 
suggesting a differential evolution between \emph{r-}process elements and Fe. This possibility is consistent with our next 
finding, namely the differential evolution of \emph{r-} and $\alpha$-elements that we obtain.\\
- We found that the $[r/\alpha]$ ratio tends to decrease with  metallicity, clearly indicating that supernovae 
having different properties contribute to the synthesis of \emph{r-}process elements and $\alpha$-elements with different 
efficiencies/yields. Since the observational evidence of \emph{r-}process synthesis via neutron-stars mergers, such a
trend cannot only be explained by the role of Type II supernovae.\\
In the context of the Second Gaia Data Release \citep{brown_2018}, this paper will be the object of an extension 
including individual stellar ages, and a study of the radial and vertical abundance gradients in the Milky Way disc.

\begin{acknowledgements}
We acknowledge financial support form the ANR 14-CE33-014-01. 
The spectra calculations were performed with the high-performance 
computing facility MESOCENTRE, hosted by OCA. This work has made 
use of the VALD database, operated at Uppsala 
University, the Institute of Astronomy RAS in Moscow, and the 
University of Vienna.

\end{acknowledgements}

\bibliographystyle{aa}
\bibliography{cite_r_s}

\end{document}